\documentclass[11pt]{article}
\pdfoutput=1
\usepackage{jcapmod}

\usepackage{tikz}
\usetikzlibrary{shapes.misc,positioning,arrows,arrows.meta,bending,matrix,shapes,fit,tikzmark,calc,external,patterns,topaths,decorations.pathmorphing,decorations.markings,decorations.pathreplacing}

\tikzset{>=latex} % for LaTeX arrow head

\usepackage[export]{adjustbox}
\usepackage{array}
\usepackage{shorthand}
\usepackage{mathtools}
\usepackage{booktabs}
\usepackage[english]{babel}
\usepackage{amsmath,amssymb,amsbsy,amstext, amsthm, simplewick, amsfonts}
\usepackage{hyperref}
\usepackage{graphicx}
\usepackage[small]{caption}
\usepackage{siunitx}
\usepackage{upgreek}
\usepackage{framed}
\usepackage{wrapfig}
\usepackage{multirow}
\usepackage{bbm}

\tikzset{fieldlines/.style={black,decoration={markings,mark=at position #1 with {\arrow[opacity=1]{latex}}},
                            postaction={decorate},line width=1},
         fieldlines/.default=0.55}
      
 \tikzset{fieldlines2/.style={black,decoration={markings,mark=at position #1 with {\arrow[opacity=1]{latex}}},
                            postaction={decorate},line width=0.5},
         fieldlines2/.default=0.55}   
 \tikzset{fieldlines3/.style 2 args={decoration={markings,mark=at position #1 with {\arrow[opacity=#2]{latex}}},
                            postaction={decorate},line width=1},
         fieldlines3/.default={0.55}{1}}  
\tikzset{cross/.style={cross out, draw=black, minimum size=2*(#1), inner sep=0pt, outer sep=0pt},%default radius will be 1pt. 
cross/.default={1pt}}

\newcommand{\tikzloop}[8]{
  \begin{scope}
    \draw (#1,#2) circle[radius=#3];
    \ifnum#8=1
    \foreach \i in {1,...,#4} {
      \draw ({#3*sin(360*\i/#4)+#1},{#3*cos(360*\i/#4)+#2}) node[rotate={-360*\i/#4},cross=#6,line width=#7] {};
    }
    \fi
    \foreach \i in {1,...,#4} {
      \filldraw[black] ({#3*sin(360*\i/#4+180/#4)+#1},{#3*cos(360*\i/#4+180/#4)+#2}) circle (#5);
    }
  \end{scope}
}
\newcounter{itemcounta}
\newcounter{itemcountb}
\newcounter{itemcountc}
\newcommand{\timeloop}[8]{
\setcounter{itemcountc}{0}
  \begin{scope}[xshift=#1,yshift=#2,rotate=180]

	\foreach[count=\i] \V in {#7} { \addtocounter{itemcountc}{\V}
	\foreach \j in {1,...,\V}{
	\filldraw ({#3*sin(360*(\i-1)/#4+180/#4+(\j-1-(\V-1)/2)*15)},{#3*cos(360*(\i-1)/#4+180/#4+(\j-1-(\V-1)/2)*15)}) circle(#5);}
	\draw  ({#3*sin(360*(\i-1)/#4+180/#4+((\V-1)/2)*15)},{#3*cos(360*(\i-1)/#4+180/#4+((\V-1)/2)*15)}) arc ({90-(360*(\i-1)/#4+180/#4+((\V-1)/2)*15)}:{90-(360*(\i-1)/#4+180/#4-((\V-1)/2)*15)}:#3);
	
    }

   	\foreach[count=\i] \LR in {#6} {
		\foreach[count=\k] \l in {#7} {
			\ifnum \k=\i
				\setcounter{itemcounta}{\l}
			\fi
			\ifnum \k=\numexpr\i-1 \relax 
				\setcounter{itemcountb}{\l}
			\fi
			\ifnum \k=\numexpr\i-1 +#4\relax 
				\setcounter{itemcountb}{\l}
			\fi	
			
		}
	\ifnum \LR =3 
			\draw ({#3*sin(360*(\i-2)/#4+180/#4)},{#3*cos(360*(\i-2)/#4+180/#4)}) arc ({90-(180/#4+360*(\i-2)/#4)}:{90-(180/#4+360*(\i-1)/#4}):#3);
	\fi
	\ifnum \LR=2
			\draw[densely dotted] ({#3*sin(360*(\i-2)/#4+180/#4)},{#3*cos(360*(\i-2)/#4+180/#4)}) arc ({90-(180/#4+360*(\i-2)/#4)}:{90-(180/#4+360*(\i-1)/#4}):#3);
	\fi
	\ifnum \LR<2

			\draw ({#3*sin(360*(\i-2)/#4+180/#4)},{#3*cos(360*(\i-2)/#4+180/#4)}) arc ({90-(180/#4+360*(\i-2)/#4)}:{90-(180/#4+360*(\i-1)/#4}):#3);
		\begin{scope}[xshift={(-1)^\LR*cos(-360*(\i-1)/#4)*7/24*(3pt+1pt *4.5)},yshift={(-1)^\LR*sin(-360*(\i-1)/#4)*7/24*(3pt+1pt *4.5)},rotate={-360*(\i-1)/#4}]
			\draw[opacity=0,fieldlines3={1}{#8}] (0,#3)--({(-1)^\LR*0.001},{#3});
		\end{scope}
	\fi

    }

  \end{scope}
}
\usepackage{selinput}

\usepackage{bm}
\usepackage{float}
\usepackage{geometry}
\usepackage{yfonts}
\usepackage{caption}
\usepackage{subcaption}
\usepackage{sidecap}
\usepackage{longtable}
\usepackage{anyfontsize}
\usepackage{dsfont}
\usepackage{tikz}
\usepackage{relsize}
\usepackage{tcolorbox}

\usepackage{xcolor}
\usepackage{xparse}

\usepackage{slashed}
\usepackage{simpler-wick}

\NewDocumentCommand{\colornucleus}{omme{_^}}{%
  \begingroup\colorlet{currcolor}{.}%
  \IfValueTF{#1}
   {\textcolor[#1]{#2}}
   {\textcolor{#2}}
    {%
     #3% the nucleus
     \IfValueT{#4}{_{\textcolor{currcolor}{#4}}}% subscript
     \IfValueT{#5}{^{\textcolor{currcolor}{#5}}}% superscript
    }%
  \endgroup
}

\definecolor{blue3}{RGB}{31, 119, 180}
\definecolor{red3}{RGB}{	214, 39, 40}
\definecolor{orange3}{RGB}{255, 127, 14}
\definecolor{green3}{RGB}{44, 160, 44}

\usepackage{array}
\newcolumntype{L}[1]{>{\raggedright\let\newline\\\arraybackslash\hspace{0pt}}m{#1}}
\newcolumntype{C}[1]{>{\centering\let\newline\\\arraybackslash\hspace{0pt}}m{#1}}
\newcolumntype{R}[1]{>{\raggedleft\let\newline\\\arraybackslash\hspace{0pt}}m{#1}}

\usepackage[framemethod=default]{mdframed}
\newmdenv[skipabove=7pt,
skipbelow=7pt,
rightline=false,
leftline=false,
topline=false,
bottomline=false,
backgroundcolor=gray!10,
linecolor=gray,
innerleftmargin=5pt,
innerrightmargin=5pt,
innertopmargin=5pt,
innerbottommargin=5pt,
leftmargin=0cm,
rightmargin=0cm,
linewidth=4pt]{eBox}
\newmdenv[skipabove=7pt,
skipbelow=7pt,
rightline=false,
leftline=false,
topline=false,
bottomline=false,
backgroundcolor=gray!10,
linecolor=gray,
innerleftmargin=5pt,
innerrightmargin=5pt,
innertopmargin=-5pt,
innerbottommargin=5pt,
leftmargin=0cm,
rightmargin=0cm,
linewidth=4pt]{eBox2}
\usepackage[framemethod=default]{mdframed}
\newmdenv[skipabove=7pt,
skipbelow=7pt,
rightline=true,
leftline=true,
topline=true,
bottomline=true,
backgroundcolor=gray!15,
linecolor=gray,
innerleftmargin=5pt,
innerrightmargin=5pt,
innertopmargin=5pt,
innerbottommargin=5pt,
leftmargin=0cm,
rightmargin=0cm,
linewidth=0.75pt]{eBox3}

\definecolor{Red}{RGB}{214, 39, 40}
\definecolor{Blue}{RGB} {31, 119, 180}
\definecolor{Orange}{RGB}{255, 153, 51}
\definecolor{Purple}{RGB}{178, 102, 255}
\definecolor{Green}{RGB}{44, 160, 44}

\newcommand{\green}[1]{\textcolor{green3}{#1}}
\newcommand{\red}[1]{\textcolor{red3}{#1}}

\newcommand{\orange}[1]{\textcolor{Orange}{#1}}

\definecolor{vio}{RGB}{19, 130, 164}
\definecolor{vioo}{RGB}{89, 2, 155}
\newcommand{\Comment}[1]{{}}
\definecolor{darkblue}{rgb}{0.15,0.35,0.55}
\definecolor{reddish}{rgb}{0.65, 0.2, 0.2}
\definecolor{darkgreen}{RGB}{50,150,0}
\definecolor{greyish}{rgb}{.90,.90,.90}
\definecolor{greyish2}{rgb}{.96,.96,.96}
\definecolor{greyish3}{rgb}{.37,.37,.37}
\definecolor{darkblue2}{rgb}{0.3,0.4,0.9}
\definecolor{Blue3}{RGB}{31, 119, 180}
\usepackage[linktocpage=true]{hyperref}
\hypersetup{
colorlinks=true,
citecolor=darkblue,
linkcolor=reddish,
urlcolor=darkblue,
pdfauthor={},
pdftitle={},
pdfsubject={}
}

\usepackage{colortbl}
\definecolor{lightgreen}{cmyk}{0.2, 0, 0.2, 0.2}
\definecolor{lightgray2}{cmyk}{0.1,0.1,0,0.1}
\definecolor{Red2}{RGB}{214, 39, 40}
\definecolor{Blue2}{RGB} {31, 119, 180}
\definecolor{Orange2}{RGB}{255, 127, 14}
\definecolor{Green2}{RGB}{44, 160, 44}

\makeatletter
\newlength{\apb@width}
\newcommand{\autoparbox}[2][c]{\settowidth{\apb@width}{#2}\parbox[#1]{\apb@width}{#2}}

\makeatother


\def\hs{\hskip 1pt}

\def\beq{\begin{equation}}
\def\eeq{\end{equation}}
\def\be{\begin{equation}}
\def\ee{\end{equation}}

\def\k{\vec k}

\newcommand{\ud}{{\rm d}}

\allowdisplaybreaks[1]
\begin{document}

\newgeometry{top=2cm, bottom=2cm, left=2cm, right=2cm}

\begin{titlepage}
\setcounter{page}{1} \baselineskip=15.5pt 
\thispagestyle{empty}

\begin{center}
{\fontsize{21}{18} \bf Geometry of Kinematic Flow}
\end{center}

\vskip 20pt
\begin{center}
\noindent
{\fontsize{14}{18}\selectfont 
Daniel Baumann\hs$^{1,2,3}$, Harry Goodhew\hs$^{1}$,
Austin Joyce\hs$^{4,5}$,\\[8pt] Hayden Lee\hs$^{6}$, 
Guilherme L.~Pimentel\hs$^{7}$  and Tom Westerdijk\hs$^{7}$}
\end{center}

\begin{center}
  \vskip8pt
\textit{$^1$  Leung Center for Cosmology and Particle Astrophysics,
Taipei 10617, Taiwan}

  \vskip8pt
\textit{$^2$  Center for Theoretical Physics,
National Taiwan University, Taipei 10617, Taiwan}

  \vskip8pt
\textit{$^3$ Institute of Physics, University of Amsterdam, Amsterdam, 1098 XH, The Netherlands}

\vskip 8pt
\textit{$^4$ 
Department of Astronomy and Astrophysics,
University of Chicago, Chicago, IL 60637, USA}

\vskip 8pt
\textit{$^5$ Kavli Institute for Cosmological Physics, 
University of Chicago, Chicago, IL 60637, USA}

\vskip 8pt
\textit{$^6$ Department of Physics and Astronomy, University of Pennsylvania, Philadelphia, PA 19104, USA}

\vskip 8pt
\textit{$^7$ Scuola Normale Superiore and INFN, Piazza dei Cavalieri 7, 56126, Pisa, Italy}
\end{center}

%=========================================
\vspace{0.4cm}
\begin{center}{\bf Abstract}
\end{center}
\noindent
We uncover a geometric organization of the differential equations for the wavefunction coefficients of conformally coupled scalars in power-law cosmologies.
To do this, we introduce a basis of functions inspired by a decomposition of the wavefunction into time-ordered components. Representing these basis functions and their singularities by graph tubings, we show that a remarkably simple rule for the merger of tubes produces the differential equations for arbitrary tree graphs (and loop integrands).  We find that the basis functions can be assigned to the vertices, edges, and facets of convex geometries (in the simplest cases, collections of hypercubes) which capture the compatibility of mergers and define how the basis functions are coupled in the differential equations.  
This organization of functions also simplifies solving the differential equations.
The merger of tubes is shown to reflect the causal properties of bulk physics, in particular the collapse of time-ordered propagators.
Taken together, these observations demystify the origin of the  {\it kinematic flow} observed in these equations~\cite{Arkani-Hamed:2023kig}.

\end{titlepage}
\restoregeometry

\newpage
\setcounter{tocdepth}{3}
\setcounter{page}{2}

\linespread{1.2}
\tableofcontents
\linespread{1.1}

\newpage
\section{Introduction}

Measurements of the cosmic microwave background have revealed that the hot Big Bang was not the beginning of time, but only the end of an earlier high-energy period~\cite{WMAP:2003syu,Spergel:1997vq,Dodelson:2003ip}. In order to learn about physics before the hot Big Bang, we study patterns in the distribution of matter in the late universe. 
In particular, we look for signatures of the pre-Big Bang evolution in the spatial correlations of large-scale density fluctuations.
In doing this, we are exploiting the fact that events that occur at different moments in time leave imprints on different scales. By 
studying correlations as a function of scale, we therefore hope to extract information about the time evolution of the primordial universe.
 
 \vskip 4pt
 The cosmological bootstrap~\cite{Arkani-Hamed:2018kmz, Baumann:2022jpr} seeks to turn this perspective around.
 Instead of following bulk time evolution to deduce the pattern of correlations, the goal is to reconstruct the space of possible correlations by imposing 
 physical consistency conditions on the reheating surface
 (i.e.~the future boundary of inflation or the past boundary of the hot Big Bang).\footnote{In
 recent years, there has been substantial progress in this endeavor; see e.g.~\cite{Arkani-Hamed:2015bza,Benincasa:2018ssx,Baumann:2019oyu,Hillman:2019wgh,Sleight:2019mgd,Sleight:2019hfp,Sleight:2020obc,Pajer:2020wxk,Albayrak:2020fyp,Armstrong:2020woi,Green:2020whw,Jazayeri:2021fvk,Salcedo:2022aal,DuasoPueyo:2023kyh,Meltzer:2020qbr,Meltzer:2021zin,DiPietro:2021sjt,Hogervorst:2021uvp,Pimentel:2022fsc,Jazayeri:2022kjy,Loparco:2023rug,SalehiVaziri:2024joi,Armstrong:2022vgl,Benincasa:2019vqr,Benincasa:2020aoj,Albayrak:2023hie}.} 
These consistency conditions are formulated as constraints on the functional form of boundary correlators, which are defined on 
the fixed time slice at the hot Big Bang.
At tree level, correlators have predictable kinematic singularities, on which they factorize~\cite{Arkani-Hamed:2017fdk,Baumann:2020dch,Goodhew:2020hob,Goodhew:2021oqg,Melville:2021lst,Baumann:2021fxj}, with coefficients related to scattering amplitudes~\cite{Raju:2012zr,Maldacena:2011nz}.
Time evolution of the bulk fields can be translated into differential equations that constrain the momentum dependence of these boundary correlators.\footnote{In de Sitter space, these differential equations can be understood as Ward identities for bulk spacetime symmetries~\cite{Maldacena:2011nz,Creminelli:2011mw,Kehagias:2012pd,Mata:2012bx,Arkani-Hamed:2018kmz,Baumann:2019oyu}, forging a close connection to  conformal field theory in momentum space~\cite{Bzowski:2013sza,Bzowski:2015pba,Bzowski:2017poo,Farrow:2018yni,Bzowski:2019kwd,Bzowski:2020kfw,Dymarsky:2014zja,Gillioz:2018mto,Gillioz:2019lgs,Gillioz:2020mdd}.}  
Solving the differential equations (with boundary conditions given by the known energy singularities), we obtain correlators for arbitrary kinematic configurations. 
 
\vskip 4pt
Recently, these differential equations were studied for the correlators of  conformally coupled scalars in a power-law cosmology~\cite{Arkani-Hamed:2023kig}. 
It was found that simple rules---called {\it kinematic flow}---predict the differential equations for arbitrary tree graphs (and loop integrands~\cite{Baumann:2024mvm}).  However, the rules were somewhat intricate, and their precise origin remained mysterious. In this paper, we introduce an alternative basis (see also~\cite{He:2024olr,De:2024zic,Fevola:2024nzj,Glew:2025ugf}) in which the kinematic flow simplifies dramatically, and  the geometric origin of the discovered pattern is revealed. 

\vskip 4pt
To describe this further, we first recall that
the bulk perturbation theory of  $n$-point correlators (and the associated wavefunction coefficients) can be represented by
Feynman graphs of the form:
 \begin{equation}
\includegraphics[valign=c]{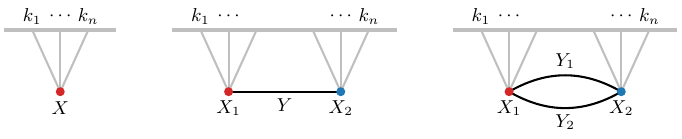}
\label{equ:Graphs}
\end{equation}
Since energy is not conserved in cosmology, each bulk interaction vertex must be integrated over time, which leads to nontrivial Feynman integrals even at tree level.
For conformally coupled fields, the resulting functions only depend on the sum of the external energies entering each vertex, which we denote by $X_v$, and the internal energies associated to the edges of the graph, which we denote by~$Y_e$. We will therefore often work with truncated graphs where the external lines are amputated, and the kinematic information is encoded in the energies $X_v$ and $Y_e$.

\vskip 4pt
Abstractly, the wavefunction coefficients generated by evaluating these Feynman diagrams are functions of the $X$ and $Y$ variables. In general, these functions can be complicated, which reflects the challenge of actually doing the integrals over time. Fortunately, much like for
loop amplitudes, the wavefunction coefficients here are part of a finite-dimensional basis of master integrals. We can discover these basis functions by taking derivatives of the original wavefunction $\psi(X_v, Y_e)$ with respect to the kinematic variables $Z_I \equiv (X_v,Y_e)$, which leads to new functions $F_i$. We repeat the process until the system of equations closes.  Arranging these functions into a finite-dimensional vector $\vec{I} \equiv [\psi, \vec{F}\hs]^T$, leads to
a differential equation of the form
\beq
{\rm d} \vec{I} = A \vec{I}\,,
\label{equ:DE}
\eeq
where ${\rm d} \equiv \sum {\rm d}Z_I \partial_{Z_I}$ and $A$ is a connection matrix. Explicitly, the connection matrix
can be written as a sum of dlog-forms $A=\sum_i A_i\, {\rm d}\log \Phi_i(Z)$, where $A_i$ are constant matrices and the functions $\Phi_i(Z)$ are called {\it letters}. The alphabet of letters captures the possible singularities of the differential equation and the associated master integrals.

\vskip 4pt
In~\cite{Arkani-Hamed:2023kig}, equation~\eqref{equ:DE} and the corresponding $A$-matrices were computed
for a number of examples (see also~\cite{De:2023xue, Chen:2023iix,Chen:2024glu,Benincasa:2024ptf,Fevola:2024nzj, Hang:2024xas,Grimm:2024tbg,Grimm:2024mbw, Grimm:2025zhv,Fan:2024iek,Liu:2024str} for related work).
Although the resulting equations at first glance look random and complicated, an intriguing pattern emerged after representing the 
letters and basis functions  by graph tubings.
Relatively simple ``merger" and ``absorption" rules predicted the differential equations for arbitrary tree graphs (and loop integrands~\cite{Baumann:2024mvm}).

\vskip 4pt
In this paper, we explain the observed patterns. In particular, we will  show that the structure of the equations further simplifies when we choose an alternative function basis.  The new basis is inspired by the structure of the bulk-to-bulk propagator, which can be separated into three independent pieces: a time-ordered piece, an anti-time-ordered piece, and a non-time-ordered piece.\footnote{This basis has previously been studied in the context of the flat-space wavefunction in~\cite{Fevola:2024nzj}, where it was shown that it splits the wavefunction into partial fractions, and in~\cite{Glew:2025ugf}, where the time orderings arose naturally from graph tubings. The differential equations were also derived in this basis in~\cite{He:2024olr}. Here, we discuss the basis in the context of the kinematic flow, and explore the geometric structures that emerge from these differential equations. Hints of these structures can also be seen in~\cite{De:2024zic}.}  
Graphically, we therefore represent each propagator as
 \begin{equation}
\includegraphics[valign=c]{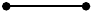} \ \ \ =\ \  \ 
\includegraphics[valign=c]{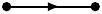}
\ \ \ +\ \  \ 
\includegraphics[valign=c]{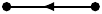}
\ \ \ -\ \  \ 
\includegraphics[valign=c]{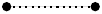}
\end{equation}
Using this structure, a graph with $e$ edges naturally separates into $3^e$ terms. For example, the wavefunction for the three-site chain can be written as the sum
\beq
\begin{aligned}
\psi \ =\ &\Big( \psi_{\includegraphics{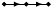}}
\ +\
 \psi_{\includegraphics{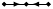}} \ +\
 \psi_{\includegraphics{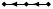}} \ + \ 
 \psi_{\includegraphics{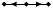}} \Big) \\
 &\ - \
\Big( \psi_{\includegraphics{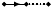}} \ +\   \psi_{\includegraphics{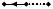}}\Big)
\ - \
\Big( \psi_{\includegraphics{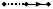}} \ +\   \psi_{\includegraphics{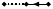}}\Big) \ + \ \psi_{\includegraphics{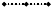}} 
 \ .
\end{aligned}
\label{equ:3site-psi}
\eeq
We construct a new basis starting from these $3^e$ functions $\psi_i$, which we can arrange into a vector~$\vec{\psi}$. Taking derivatives with respect to the kinematic variables $Z_I \equiv (X_v,Y_e)$ now leads to new functions~$G_i$, and the vector of master integrals is $\vec I \equiv [\vec{\psi},\vec{G}\hs]^T$. 
The additional basis functions $G_i$ are related to the ``collapse" of time-ordered propagators, and we end up with a basis involving $4^e$ functions.
The notion of collapsing propagators defines a partial ordering, so that these basis functions can be assigned to the vertices, edges, facets, etc.~of convex geometries (in the simplest cases, collections of hypercubes) which capture the compatibility of different ways to collapse propagators and define how the basis functions are coupled in the differential equations.
For example, in the case of the three-site chain, the $16$ basis functions can be arranged geometrically as
\beq
\raisebox{-4.9em}{ \begin{tikzpicture}[line width=0.75 pt, scale=0.65]
 \begin{scope}[xshift=-5cm]
\node at (0,0.5)  {\scalebox{1.5}{\includegraphics{Figures/Subscripts/psi00.pdf}}};
\draw[fill=Blue,Blue] (0,0) circle (5pt);
\end{scope}
\begin{scope}[yshift=1.5cm]
\draw[fill=Blue,Blue] (-1.5,0) circle (5pt);
\draw[fill=Blue,Blue] (1.5,0) circle (5pt);
\draw[Blue,line width=2pt] (-1.5,0) -- (1.5,0);
\node at (-1.5,0.5)  {\scalebox{1.5}{\includegraphics{Figures/Subscripts/psi+0.pdf}}};
\node at (1.5,0.5)  {\scalebox{1.5}{\includegraphics{Figures/Subscripts/psi-0.pdf}}};
\node at (0,-0.4)  {\scalebox{1.5}{\includegraphics{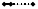}}};
\end{scope}
\begin{scope}[yshift=-1.5cm]
\draw[fill=Blue,Blue] (-1.5,0) circle (5pt);
\draw[fill=Blue,Blue] (1.5,0) circle (5pt);
\draw[Blue,line width=2pt] (-1.5,0) -- (1.5,0);
\node at (-1.5,0.5)  {\scalebox{1.5}{\includegraphics{Figures/Subscripts/psi0+.pdf}}};
\node at (1.5,0.5)  {\scalebox{1.5}{\includegraphics{Figures/Subscripts/psi0-.pdf}}};
\node at (0,-0.4)  {\scalebox{1.5}{\includegraphics{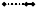}}};
\end{scope}
\begin{scope}[xshift=8cm]
\draw[fill=Blue,Blue] (-2,2) circle (5pt);
\draw[fill=Blue,Blue] (2,2) circle (5pt);
\draw[fill=Blue,Blue] (-2,-2) circle (5pt);
\draw[fill=Blue,Blue] (2,-2) circle (5pt);
\draw[fill=Blue, opacity=0.15] (-2,2) -- (2,2) -- (2,-2) -- (-2,-2) --cycle;
\draw[Blue,line width=2pt] (-2,2) -- (2,2) -- (2,-2) -- (-2,-2) --cycle;
\node at (0,0)  {\scalebox{1.5}{\includegraphics{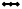}}};
\node at (-2,2.5)  {\scalebox{1.5}{\includegraphics{Figures/Subscripts/psi++.pdf}}};
\node at (2,2.5)  {\scalebox{1.5}{\includegraphics{Figures/Subscripts/psi+-.pdf}}};
\node at (0,1.6)  {\scalebox{1.5}{\includegraphics{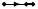}}};
\node at (-2,-2.5)  {\scalebox{1.5}{\includegraphics{Figures/Subscripts/psi-+.pdf}}};
\node at (2,-2.5)  {\scalebox{1.5}{\includegraphics{Figures/Subscripts/psi--.pdf}}};
\node at (0,-1.6)  {\scalebox{1.5}{\includegraphics{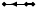}}};
\node at (-3,0)  {\scalebox{1.5}{\includegraphics{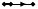}}};
\node at (3.,0)  {\scalebox{1.5}{\includegraphics{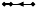}}};
\end{scope}
\end{tikzpicture}}
\label{equ:3-geo}
\eeq
The functions are separated into disjoint sectors graded by the number of non-time-ordered propagators (dashed lines). The differential equations don't mix functions between the different sectors.  Each vertex gets assigned one of the nine contributions to the wavefunction shown in~\eqref{equ:3site-psi}.  The ``edge functions"~$G_i$ then arise from collapsing one of the time-ordered propagators in the attached 
``vertex functions''~$\psi_i$. Similarly, the function on the face of the square, $G_{\includegraphics{Figures/Subscripts/psicc.pdf}}$,  is obtained by collapsing the remaining time-ordered propagator in the edge functions $G_i$. This fully captures the way the different basis functions are connected in the differential equations. Edge functions appear as sources in the differential equations of the attached vertex functions, while the face function becomes a source in the equations for the related edge functions. This predicts the $A$-matrix in~\eqref{equ:DE}.

\vskip 4pt
As in~\cite{Arkani-Hamed:2023kig}, we introduce a graphical representation of the letters and functions in terms of graph tubings. The tubings describing the letters  are the same as~\cite{Arkani-Hamed:2023kig}, while the tubings representing functions will have a slightly different meaning, because we are using a different basis. There are $4^e$  complete tubings of the (marked) graph, which are in one-to-one correspondence with the $4^e$ basis functions introduced above. The graph tubings, however, also have an autonomous definition in terms of the boundary kinematics without any reference to the bulk dynamics. We further find that the collapse of the time-ordered propagators in the bulk description corresponds to the merger of adjacent tubes on the graph tubings. The compatibility of such mergers explains how the basis functions are related in the differential equations.
From this perspective, the rules of the kinematic flow simplify drastically into a single ``merger" rule. Repeated application of this rule produces the differential equations for arbitrary tree graphs (and loop integrands).   Integrability of the differential equations is also transparent in the new basis, and the geometric organization of the equations simplifies the procedure of actually solving them.

\paragraph{Outline} The outline of this paper is as follows: In Section~\ref{sec:background}, we review the toy model of conformally coupled scalars in a power-law cosmology, and explain how to compute wavefunction coefficients in this theory. Experts (and readers of~\cite{Arkani-Hamed:2023kig}) can skip this part.
In Section~\ref{sec:BulkTime}, we study the time integrals arising in the perturbative computations of the wavefunction. We identify the geometrical pattern underlying the differential equations for these integrals.
In Section~\ref{sec:boundary}, we present a more boundary-centric description of the same physics. Each basis function is represented by a graph tubing and we define associated energy integrals in the boundary kinematic space. We then show that the differential equations for these energy integrals are predicted by an extremely simple kinematic flow rule. 
Finally, our conclusions are stated in Section~\ref{sec:conclusions}.

\vskip 4pt 
Two appendices contain additional material: In Appendix~\ref{app:solving}, we explicitly solve the differential equations for the two-site chain, which illustrates the implementation of boundary conditions discussed abstractly in Section~\ref{ssec:BC}. In~Appendix~\ref{app:trp3}, we provide a brief discussions of the equations for ${\rm tr}\hs \phi^3$ theory, where the result depends on the sum of Feynman graphs in different exchange channels. We show that some basis functions are shared between the different channels, reducing the naive size of the basis, and explain how this information can be represented geometrically.

%%%%%%%%%%%%%%%%%%%%%%%%%%%%%%%%%%%%%%%%%%%%%%%%%%%%%%%
\section{Setup and Background}
\label{sec:background}

The theory of interest (as in~\cite{Arkani-Hamed:2023kig}) is a conformally coupled scalar field in a power-law FRW universe, with polynomial interactions:\footnote{Here, we specialize to $D=4$ spacetime dimensions. It is straightforward to generalize this action to a generic spacetime dimension $D$, which is necessary when implementing dimensional regularization of loop integrals.}
\beq
  S= \int {\rm d}^4 x\, \sqrt{-g}\left[-\frac{1}{2}(\partial \phi)^2-\frac{1}{12}R\phi^2-  \sum_{p=3}^\infty \frac{\lambda_p}{p!} \phi^p\right] ,
  \label{equ:action}
\eeq
where $R$ is the Ricci scalar.  Although this is a toy model for cosmological correlations, it shares many features with more realistic models and is simple enough to study in great detail. In particular, if the universe evolves as a power law in conformal time, then the action~\eqref{equ:action} can be transformed into the action of a field in flat space with time-dependent couplings, which greatly simplifies the computation of  
correlation functions (or wavefunction coefficients) in this theory.

\subsection{Power-Law Cosmology}

A universe that is dominated by a single fluid with a constant equation of state will evolve as a power law in conformal time. In this case, the spacetime metric takes the form
\beq
{\rm d}s^2 = a^2(\eta) (-{\rm d}\eta^2 + {\rm d} {\bf x}^2)\,, \qquad\quad a(\eta) \equiv \left(\frac{\eta}{\eta_0}\right)^{-(1+\varepsilon)}\,,
\label{equ:metric}
\eeq
where 
 $\varepsilon$ and $\eta_0$ are constant parameters.
  We will take the conformal time $\eta$ to run from $-\infty$ to $0$, and choose units such that $\eta_0 \equiv - 1$.
 Many cosmologies of interest fit into this family, 
such as $\varepsilon = 0$ (de Sitter), $\varepsilon \approx 0$ (inflation), $\varepsilon =-1$ (Minkowski), $\varepsilon =-2$ (radiation) and $\varepsilon=-3$ (matter).  
With this parameterization, the universe experiences accelerated expansion
  for $\varepsilon > -1$ and decelerated collapse for $\varepsilon < -1$.
We will be interested in correlations at a fixed time $\eta_* = 0$, which for an accelerating universe is the future boundary of the spacetime. We will impose Bunch--Davies initial conditions in the far past, $\eta \to -\infty$, although the differential equations governing the correlators don't depend on this assumption and only their boundary conditions would be affected by a different choice of initial conditions. 
 
\vskip 4pt
We now consider conformally coupled scalars in this background. Under a Weyl transformation $g_{\mu \nu} \mapsto a^{2} g_{\mu \nu}$, $\phi \mapsto a^{-1}\phi$, the action (\ref{equ:action}) becomes that of a massless field in flat space with time-dependent couplings:
\beq
S = \int {\rm d}^4 x \left[-\frac{1}{2}(\partial\phi)^2 - \sum_{p=3}^\infty\frac{\lambda_p(\eta)}{p!}\phi^p\right] ,
\eeq
where $\lambda_p(\eta) = \lambda_p\hs a(\eta)^{4-p}$.
Specializing to the power-law scaling in (\ref{equ:metric}), we have
\beq
\lambda_p(\eta) = \lambda_p (-\eta)^{-(1+\alpha_p)}\,, \quad {\rm with} \quad \alpha_p \equiv (4-p)(1+ \varepsilon) -1\,.
\eeq
For a cubic interaction, $\alpha_3 = \varepsilon$, while for a quartic interaction the theory is conformally invariant and $\alpha_4=-1$, so that the coupling is time-independent as in flat space. All the time dependence has been absorbed into the couplings, $\lambda_p(\eta)$, so that the mode functions of the field are the same as in flat space, $\phi_k^{\rm flat}(\eta)=e^{ik\eta}/\sqrt{2k}$, which simplifies computations in this theory.

\subsection{Wavefunction Coefficients}
\label{sec:WFC}

We want to study correlation functions of the field $\phi$ evaluated on the boundary at $\eta_* = 0$.
These ``boundary correlators" can be written as a path integral
\beq
\langle \varphi({\bf x}_1) \cdots \varphi({\bf x}_N) \rangle = \int {\cal D} \varphi\, \varphi({\bf x}_1) \cdots \varphi({\bf x}_N)\, |\Psi[\varphi]|^2\,,
\eeq
where $\varphi({\bf x}) \equiv \phi(0,{\bf x})$ is the boundary value of the field. This is just the generalization of the ordinary formula in quantum mechanics, where field configurations are weighted by the square of the wavefunction(al) $\Psi[\varphi]$, which provides a probability distribution on the space of field profiles.
For small fluctuations, we expand this wavefunction (in Fourier space) as
\beq
    \Psi[\varphi]=\exp\left[-\sum_{n=2}^\infty \int \frac{\ud^3 k_1}{(2\pi)^3}\cdots \frac{\ud^3 k_n}{(2\pi)^3}\,(2\pi)^3 \delta(\textbf{k}_1+\dots+\textbf{k}_n)\,\psi_n(\textbf{k}_1,\cdots,\textbf{k}_n) \,\varphi_{\textbf{k}_1}\cdots \varphi_{\textbf{k}_n}\right] ,
\eeq
where the kernel functions $\psi_n$ are called {\it wavefunction coefficients}.  These wavefunction coefficients are the objects that we are interested in computing.
In perturbation theory, they carry the same information as the boundary correlators.

\vskip4pt
Each wavefunction coefficient can be represented by a sum over Feynman graphs, like those shown in \eqref{equ:Graphs}.  In general, the kinematic data that these wavefunction coefficients depend on is a set of $n$ three-vectors, $\textbf{k}_a$, subject to momentum conservation. (In a small abuse of terminology, we will refer to the lengths of these vectors $k_a \equiv \lvert {\bf k}_a\rvert$ as ``energies".)
For conformally coupled fields, a simplification occurs because the result only depends on the sum of the external energies entering each vertex, which we denote by $X_a$, and the internal energies associated to the edges of the graph, which we denote by $Y_b$. 
We will therefore often work with truncated graphs, where the external lines are amputated and an energy $X_a$ is assigned to each vertex.

\vskip 4pt
\newpage
Operationally,  wavefunction coefficients are computed using the following {\it Feynman rules}:\footnote{The precise form of the propagators entering the Feynman rules depends on the choice of initial conditions. We are assuming that fields begin in the adiabatic vacuum, so that we are computing the analogue of the Bunch--Davies state for the interacting theory~\cite{Arkani-Hamed:2015bza,Arkani-Hamed:2018kmz,Green:2020whw}. The differential equations satisfied by $\psi_n$ do not depend on this choice; different vacuum choices correspond to imposing different boundary conditions~\cite{Arkani-Hamed:2018kmz,Arkani-Hamed:2023kig} (see Section~\ref{ssec:BC}).}
\begin{itemize}
\item {\it Bulk-to-boundary:} We assign a bulk-to-boundary propagator to every external line:
\beq
K(k,\eta) = e^{ik\eta}\,,
\label{eq:bbdyprop}
\eeq
where $k = |\textbf{k}|$.  At each vertex, the propagators of the external lines entering the vertex combine into $e^{iX_v\eta}$, where  $X_v$ is the sum of the energies $k_a$ flowing into that vertex.

\item {\it Bulk-to-bulk:} We assign a bulk-to-bulk propagator to every internal line:
\beq
G(Y;\eta,\eta') = \frac{1}{2Y} \left[e^{-iY(\eta'-\eta)} \theta(\eta'-\eta)+ e^{-iY(\eta-\eta')} \theta(\eta-\eta')  - e^{iY(\eta+\eta')}\right] .\label{eq:Gbb}
\eeq
This propagator has three pieces: 1) A time-ordered piece, 2) an anti-time-ordered piece, and 3) a non-time-ordered piece.  The sum of the two time-ordered contributions is the Feynman propagator, while the extra non-time-ordered piece is required by the boundary condition $G(Y;\eta,0) = 0$.

It is convenient to work with a rescaled bulk-to-bulk propagator, $\hat G \equiv 2Y G$, which will compute a rescaled wavefunction coefficient $\hat \psi_n$. These wavefunction coefficients differ from the true wavefunction coefficients by a factor of $2Y$ for each internal line.
Below (as in~\cite{Arkani-Hamed:2023kig}), all wavefunction coefficients will be of this rescaled form, but it is straightforward to divide by these internal energy factors to obtain the true wavefunction coefficients. We will drop the hat to avoid clutter.

\item {\it Vertex integrals:} To each vertex, we assign a (time-dependent) coupling constant $i\lambda_p(\eta_a)$ and we integrate over the time $\eta_a$ at which the interaction takes place. 

\item {\it Loop integrals:} We integrate over any unfixed loop momenta.
\end{itemize}
Using these rules, it is straightforward to write down time integrals for any Feynman graph. However, computing these integrals can be challenging, particularly as graphs get more complex. We address this challenge in the next section. 

%%%%%%%%%%%%%%%%%%%%%%%%%%%%%%%%%%%%%%%%%%%%%%%%%%%%%%%%%%%%%
\newpage
\section{Bulk Time Integrals}
\label{sec:BulkTime}

We wish to study the structure of the time integrals that compute wavefunction coefficients for the conformally coupled scalar. 
It will be convenient to use the following
diagrammatic representation of the bulk-to-bulk propagator~\eqref{eq:Gbb}, which manifests the fact that it is built from three pieces: 
 \begin{equation}
\includegraphics[valign=c]{Figures/Equations/2trimmed.pdf} \ \ \ =\ \  \ 
\includegraphics[valign=c]{Figures/Equations/2timeright.pdf}
\ \ \ +\ \  \ 
\includegraphics[valign=c]{Figures/Equations/2timeleft.pdf}
\ \ \ -\ \  \ 
\includegraphics[valign=c]{Figures/Equations/2disconnected.pdf} \ \ ,
\end{equation}
where the arrows on the internal line capture the time ordering of the vertices involved.
We will use this splitting of the propagator to define a natural set of basis functions associated to directed graphs.
These functions form a finite-dimensional vector space, and we will derive the coupled differential equations that they satisfy.\footnote{These differential equations were previously derived in~\cite{He:2024olr}, who also gave them a graphical interpretation. Here, we wish to emphasize the partial order and compatibility relations between functions that makes it possible to organize them into a geometry, along with the combinatorial interpretation of the differential equations.} 
We will discover that these differential equations obey an interesting pattern.

\subsection{Differential Equations}
\label{ssec:DE}

To see the structure underlying the differential equations satisfied by the wavefunction,
it will be instructive to start with a few explicit examples. The patterns will very quickly become apparent, and the generalization to arbitrary tree graphs (and even loop integrands) will be straightforward.

\paragraph{Two-site chain}
The simplest nontrivial example is the two-site chain, corresponding to a single bulk exchange, with an arbitrary number of external lines. It is convenient to separate the wavefunction into three pieces (matching the three parts of the bulk-to-bulk propagator):
\beq
\psi \ =\ \psi_{\includegraphics{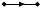}}
\ +\
 \psi_{\includegraphics{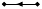}}
\ - \
\psi_{\includegraphics{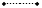}}\ ,
\eeq
where two of the components have definite time orderings and one is un-time-ordered. Using the Feynman rules articulated in Section~\ref{sec:WFC}, we write the time integrals associated to each piece as
\begin{align}
\psi_{\includegraphics{Figures/Subscripts/psi+.pdf}} &\ \equiv \ N \int \frac{{\rm d} \eta_1 \hs {\rm d} \eta_2}{(-\eta_1)^{1+\alpha_1} (-\eta_2)^{1+\alpha_2}} \,e^{i(X_1+Y)\eta_1} e^{i(X_2-Y)\eta_2} \theta(\eta_2-\eta_1) \,, \label{equ:psiT}\\
\psi_{\includegraphics{Figures/Subscripts/psi-.pdf}}  &\ \equiv \ N \int \frac{{\rm d} \eta_1 \hs {\rm d} \eta_2}{(-\eta)^{1+\alpha_1} (-\eta_2)^{1+\alpha_2}} \,e^{i(X_1-Y)\eta_1} e^{i(X_2+Y)\eta_2} \theta(\eta_1-\eta_2) \,, \label{equ:psiTT}\\
\psi_{\includegraphics{Figures/Subscripts/psi0.pdf}}  &\ \equiv \ N \int \frac{{\rm d} \eta_1 \hs {\rm d} \eta_2}{(-\eta_1)^{1+\alpha_1} (-\eta_2)^{1+\alpha_2}} \,e^{i(X_1+Y)\eta_1} e^{i(X_2+Y)\eta_2}\,, \label{equ:psiD}
\end{align}
where $N$ is an (irrelevant) overall normalization factor which depends on the precise couplings participating in the interaction. Recall that we are using the rescaled bulk-to-bulk propagator, $\hat G = 2Y G$, to compute these integrals, which explains the absence of overall factors of $(2Y)^{-1}$. The benefit of this rescaling is that now the kinematic variable $Y$ only enters these expressions in combination with $X_1$ or $X_2$.\footnote{This is an important difference between the individual time-ordered contributions and the full wavefunction. The (non-rescaled) wavefunction does not have a singularity at $Y=0$, even though the individual contributions do. In the case at hand, the additional singularities do not lead to much complication because we can multiply by appropriate factors of the $Y$'s to compute a rescaled object, and then divide by the factors at the end.} In principle, the vertex factors can be different for the two vertices, leading to different values of $\alpha_a$, which we will keep track of. Finally, note also that $\psi_{\includegraphics{Figures/Subscripts/psi+.pdf}}$ and $\psi_{\includegraphics{Figures/Subscripts/psi-.pdf}}$ are related by interchanging $X_1\leftrightarrow X_2$ and $\alpha_1\leftrightarrow \alpha_2$.

\vskip4pt
We now derive the first-order differential equations satisfied by these basis functions.
We will demonstrate the following two facts:
1) The differential of the non-time-ordered piece~\eqref{equ:psiD} is proportional to itself. 
 2) The differentials of the time-ordered pieces~\eqref{equ:psiT} and~\eqref{equ:psiTT} involve a new source function:
 \beq
F_{\includegraphics{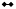}}  \equiv N \int \frac{{\rm d} \eta}{(-\eta)^{1+\alpha_1+\alpha_2}}\, e^{i(X_1+X_2)\eta}\,.
\label{equ:psi-c}
\eeq
This source function is obtained by ``collapsing" the propagator connecting the two vertices. The collapse of the propagator in this way gives a physical interpretation to the ``contracted diagrams" considered in~\cite{Grimm:2025zhv}.

\vskip4pt
As we show in the box below, the resulting differential equations are
\begin{align}
{\rm d}  \psi_{\includegraphics{Figures/Subscripts/psi0.pdf}}   &\, = \, \Big(
\alpha_1   \, {\rm d} \log(X_1+Y)   \, +\, \alpha_2   \, {\rm d} \log(X_2+Y)   \Big) \, \psi_{\includegraphics{Figures/Subscripts/psi0.pdf}}\ ,  \label{equ:dpsi0}\\
{\rm d}\psi_{\includegraphics{Figures/Subscripts/psi+.pdf}}  &\, = \, \Big(
\alpha_1\, \psi_{\includegraphics{Figures/Subscripts/psi+.pdf}}   \, - \, F_{\includegraphics{Figures/Subscripts/psic.pdf}}  \Big)   \, {\rm d} \log(X_1+Y)   \, +\, \Big(\alpha_2 \,\psi_{\includegraphics{Figures/Subscripts/psi+.pdf}}  \, + \, F_{\includegraphics{Figures/Subscripts/psic.pdf}}   \Big)   \, {\rm d} \log(X_2-Y) \,, \label{equ:dpsi+}\\
{\rm d}\psi_{\includegraphics{Figures/Subscripts/psi-.pdf}}  &\, = \, \Big(
\alpha_1\, \psi_{\includegraphics{Figures/Subscripts/psi-.pdf}}   \, + \, F_{\includegraphics{Figures/Subscripts/psic.pdf}}  \Big)   \, {\rm d} \log(X_1-Y)   \, +\, \Big(\alpha_2\,\psi_{\includegraphics{Figures/Subscripts/psi-.pdf}}  \, - \, F_{\includegraphics{Figures/Subscripts/psic.pdf}}   \Big)   \, {\rm d} \log(X_2+Y) \,, \label{equ:dpsi-}\\[2pt]
{\rm d} 
F_{\includegraphics{Figures/Subscripts/psic.pdf}}  &\, = \, (\alpha_1 + \alpha_2)\, 
F_{\includegraphics{Figures/Subscripts/psic.pdf}} \, {\rm d} \log(X_1+X_2)\,. \label{equ:dpsic}
\end{align}
We see that the two time-ordered contributions are sourced by the function with the collapsed propagator, while 
this source function and the disconnected contribution close on themselves. It is important to note that even though we were forced to introduce a new function $F_{\includegraphics{Figures/Subscripts/psic.pdf}} $ into the differential system, only the functions $\psi_{\includegraphics{Figures/Subscripts/psi0.pdf}}$, $\psi_{\includegraphics{Figures/Subscripts/psi+.pdf}}$ and $\psi_{\includegraphics{Figures/Subscripts/psi-.pdf}}$ appear in the wavefunction.

\vskip 4pt
\begin{eBox3}
{\small {\bf Derivation:} In the following, we derive the differential equations~\eqref{equ:dpsi0}--\eqref{equ:dpsic}:
\begin{itemize} 
\item First, we consider the $X_1$-derivative of the disconnected contribution~\eqref{equ:psiD}: 
 \begin{align}
 \partial_{X_1} 
\psi_{\includegraphics{Figures/Subscripts/psi0.pdf}}   &\ = \ N \int \frac{{\rm d} \eta_1 \hs {\rm d} \eta_2}{(-\eta_1)^{1+\alpha_1} (-\eta_2)^{1+\alpha_2}} \,\Big[\partial_{X_1} e^{i(X_1+Y)\eta_1}\Big] e^{i(X_2+Y)\eta_2} \nonumber \\
&\ = \ N \int \frac{{\rm d} \eta_1 \hs {\rm d} \eta_2}{(-\eta_1)^{1+\alpha_1} (-\eta_2)^{1+\alpha_2}} \, \left[\frac{\eta_1}{X_1+Y}\partial_{\eta_1}e^{i(X_1+Y)\eta_1}\right] e^{i(X_2+Y)\eta_2}  \nonumber \\
&\ = \ \frac{N}{X_1+Y}  \int \frac{{\rm d} \eta_1 \hs {\rm d} \eta_2}{(-\eta_2)^{1+\alpha_2}} \Big[\partial_{\eta_1} (-\eta_1)^{-\alpha_1} \Big] e^{i(X_1+Y)\eta_1}  e^{i(X_2+Y)\eta_2} \nonumber \\
&\ = \ \frac{\alpha_1}{X_1+Y} \
  \psi_{\includegraphics{Figures/Subscripts/psi0.pdf}}  \ .
\end{align}
 In the second line, we traded the $X_1$-derivative for an $\eta_1$-derivative using the relation between derivatives, $x\hs \partial_x e^{ix\eta} = \eta\hs \partial_\eta e^{ix\eta}$, and then integrated by parts in the third line. This generates a singularity at $X_1+Y=0$, with a coefficient that is $\alpha_1$ times the original function $\psi_{\includegraphics{Figures/Subscripts/psi0.pdf}}$\ . Following the same steps, the $X_2$-derivative is
\beq
 \partial_{X_2} 
\psi_{\includegraphics{Figures/Subscripts/psi0.pdf}}  
\ = \ \frac{\alpha_2}{X_2+Y} \
 \psi_{\includegraphics{Figures/Subscripts/psi0.pdf}}  \ .
\eeq
The derivative with respect to the internal energy, $\partial_Y \psi_{\includegraphics{Figures/Subscripts/psi0.pdf}}$, can be inferred from the derivatives with respect to the external energies $X_1$ and $X_2$, and therefore doesn't have to be considered separately. Hence, we conclude   
that the total differential is~\eqref{equ:dpsi0}.

 \item Next, we look at the $X_1$-derivative of the time-ordered contribution~\eqref{equ:psiT}:
  \begin{align}
 \partial_{X_1} 
 \psi_{\includegraphics{Figures/Subscripts/psi+.pdf}}    &\ = \ N \int \frac{{\rm d} \eta_1 \hs {\rm d} \eta_2}{(-\eta_1)^{1+\alpha_1} (-\eta_2)^{1+\alpha_2}} \,\Big[\partial_{X_1}e^{i(X_1+Y)\eta_1}\Big] e^{i(X_2-Y)\eta_2} \theta(\eta_2-\eta_1) \nonumber \\
&\ = \ N \int \frac{{\rm d} \eta_1 \hs {\rm d} \eta_2}{(-\eta_1)^{1+\alpha_1} (-\eta_2)^{1+\alpha_2}} \, \left[\frac{\eta_1}{X_1+Y}\partial_{\eta_1}e^{i(X_1+Y)\eta_1}\right] e^{i(X_2-Y)\eta_2} \theta(\eta_2-\eta_1) \nonumber \\
&\ = \ \frac{N}{X_1+Y}  \int \frac{{\rm d} \eta_1 \hs {\rm d} \eta_2}{(-\eta_2)^{1+\alpha_2}} \partial_{\eta_1}\Big[ (-\eta_1)^{-\alpha_1} \theta(\eta_2-\eta_1) \Big] e^{i(X_1+Y)\eta_1}  e^{i(X_2-Y)\eta_2} \nonumber \\
&\ = \ \frac{\alpha_1}{X_1+Y} \
\psi_{\includegraphics{Figures/Subscripts/psi+.pdf}}  
\ - \ \frac{1}{X_1+Y}\,F_{\includegraphics{Figures/Subscripts/psic.pdf}}   \,,
\end{align}
where we got an extra function $F_{\includegraphics{Figures/Subscripts/psic.pdf}}$---defined in~\eqref{equ:psi-c}---from the derivative of the theta-function. 
Note that this function is obtained by the replacement $\theta(\eta_2-\eta_1) \to \eta_1 \hs \delta(\eta_2-\eta_1)$ in~\eqref{equ:psiT}.
Similarly, the $X_2$-derivative yields
   \begin{align}
 \partial_{X_2} 
\psi_{\includegraphics{Figures/Subscripts/psi+.pdf}}   &\ = \ N \int \frac{{\rm d} \eta_1 \hs {\rm d} \eta_2}{(-\eta_1)^{1+\alpha_1} (-\eta_2)^{1+\alpha_2}} \,e^{i(X_1+Y)\eta_1} \Big[\partial_{X_2} e^{i(X_2-Y)\eta_2}\Big] \theta(\eta_2-\eta_1) \nonumber \\
&\ = \ N \int \frac{{\rm d} \eta_1 \hs {\rm d} \eta_2}{(-\eta_1)^{1+\alpha_1} (-\eta_2)^{1+\alpha_2}} \, e^{i(X_1+Y)\eta_1}  \left[\frac{\eta_2}{X_2-Y}\partial_{\eta_2}e^{i(X_2-Y)\eta_2}\right] \theta(\eta_2-\eta_1) \nonumber \\
&\ = \ \frac{N}{X_2-Y}  \int \frac{{\rm d} \eta_1 \hs {\rm d} \eta_2}{(-\eta_1)^{1+\alpha_1}} \partial_{\eta_2}\Big[ (-\eta_2)^{-\alpha_2} \theta(\eta_2-\eta_1) \Big] e^{i(X_1+Y)\eta_1}  e^{i(X_2-Y)\eta_2} \nonumber \\
&\ = \ \frac{\alpha_2}{X_2-Y} \
\psi_{\includegraphics{Figures/Subscripts/psi+.pdf}} 
\ + \ \frac{1}{X_2-Y}\,F_{\includegraphics{Figures/Subscripts/psic.pdf}}  \ .
\end{align}
Together, these imply that the differential of the time-ordered contribution is~\eqref{equ:dpsi+}.

\item By symmetry, the differential of the anti-time-ordered contribution~\eqref{equ:psiTT} is~\eqref{equ:dpsi-}.

\item Finally, the derivatives of the new function $F_{\includegraphics{Figures/Subscripts/psic.pdf}}$ can be computed as
\begin{align}
\partial_{X_1}F_{\includegraphics{Figures/Subscripts/psic.pdf}} &\ =\ N \int \frac{{\rm d} \eta}{(-\eta)^{1+\alpha_1+\alpha_2}}\, \partial_{X_1}e^{i(X_1+X_2)\eta} \nonumber \\
&\ = \ N \int  \frac{{\rm d} \eta}{(-\eta)^{1+\alpha_1+\alpha_2}}\, \frac{\eta}{X_1+X_2} \partial_\eta e^{i(X_1+X_2)\eta} \nonumber \\
&\ = \ \frac{N}{X_1+X_2} \int {\rm d} \eta \, e^{i(X_1+X_2)\eta} \partial_\eta(-\eta)^{-\alpha_1 -\alpha_2} \nonumber \\[2pt]
&\ = \ \frac{\alpha_1 + \alpha_2}{X_1+X_2}\, F_{\includegraphics{Figures/Subscripts/psic.pdf}}  \ , \\[10pt]
\partial_{X_2}F_{\includegraphics{Figures/Subscripts/psic.pdf}} &\ = \ \frac{\alpha_1 + \alpha_2}{X_1+X_2}\, F_{\includegraphics{Figures/Subscripts/psic.pdf}}  \ ,
\end{align}
 and hence the differential is~\eqref{equ:dpsic}.
 \end{itemize}
}
\end{eBox3}

\vspace{0.15cm}
\paragraph{More general graphs}
The above derivation makes it clear that the equations for larger tree graphs can be obtained recursively.
First, when we add additional non-time-ordered propagators to vertices, the derivatives with respect to these new kinematic variables don't talk to the other variables and we just get extra additive terms in the differential equations like those in~\eqref{equ:dpsi0}. On the other hand, adding extra time-ordered propagators will give a product of Heaviside theta-functions. Taking derivatives of such a product of theta-functions returns a sum of delta-functions which collapse these propagators one at a time. 

\vskip4pt
We can see this more explicitly by considering a single vertex of a general (directed) graph $\Gamma$, which locally looks like
\begin{equation}\label{eq:Va}
\raisebox{-1.05cm}{\includegraphics[scale=1]{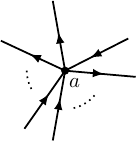}}
\end{equation}
so that the vertex $a$ is connected to the rest of the graph by $j$ time-ordered edges. The corresponding  integral over $\eta_a$ is
\be
\psi_\Gamma = \int\frac{\ud\eta_a}{(-\eta_a)^{1+\alpha_a}} \,e^{i(X_a\pm Y_{a1}\pm \cdots \pm Y_{aj})\eta_a}\, \prod_{i=1}^j  \left(\int\frac{\ud\eta_i}{(-\eta_i)^{1+\alpha_i}}\, \theta(\mp[\eta_a-\eta_i]) \right) \hat F\,,
\label{eq:Ffunction}
\ee
where $Y_{ai}$ ($i=1,\cdots, j$) are the energies carried by the internal lines, whose signs are determined by whether the time ordering points towards the vertex (bottom sign)  or away from it (top sign), and $\hat F$ denotes the rest of the integral, which does not depend on $\eta_a$. It is straightforward to derive a differential equation satisfied by $\psi_\Gamma$ from this time-integral perspective (see also~\cite{He:2024olr}). We consider the following total derivative
\be
\label{eq:IBP2}
\int\ud\eta_a\frac{\partial}{\partial\eta_a}\left[(-\eta_a)^{-\alpha_a} \,e^{i(X_a\pm Y_{a1}\pm \cdots \pm Y_{aj})\eta_a} \prod_{i=1}^j  \left(\int\frac{\ud\eta_i}{(-\eta_i)^{1+\alpha_i}}\, \theta(\mp[\eta_a-\eta_i]) \right)\hat F\,
\right] = 0\,.
\ee
By performing similar manipulations as in the two-site case, we see that this can be written in terms of $\psi_\Gamma$ as 
\be\label{eq:IBPs}
\alpha_a \psi_\Gamma -(X_a\pm Y_{a1}\pm \cdots \pm Y_{aj})\partial_{X_a}\psi_\Gamma + \sum_{i=1}^j (\mp)_i \hs F_{a\to i} = 0\,,
\ee
where the new functions $F_{a\to i}$ arise when the time derivative acts on the theta-function and turns it into a delta-function, effectively collapsing the internal line. The sign $(\mp)_i$ depends on whether the time ordering points towards $(+)$ or away from $(-)$ the vertex $i$. We obtain these source functions through the replacement $\theta(\mp[\eta_a-\eta_i]) \mapsto \eta_{i} \delta(\eta_a - \eta_i)$ in \eqref{eq:Ffunction}. Since $F_{a\to i}$ is of the same form as \eqref{eq:Ffunction}, we can continue this process until all theta-functions are 
delta-functions.

\vskip 4pt
The integration-by-parts relations~\eqref{eq:IBP2} naturally lead to equations like~\eqref{eq:IBPs} involving derivatives with respect to external energies (or the sums of energies flowing into vertices). The kinematic variables also include the internal energies, $Y_{ab}$. However, these internal variables always enter integrals in linear factors along with the external energies. (This is a consequence of our rescaling the bulk-to-bulk propagator as $\hat G = 2Y G$.) Hence, constructing the differential equation for the external energies in terms of the $\ud \log$-forms of these linear factors will always also give us the correct partial derivatives with respect to the internal energies.

\vskip4pt
We see from this discussion that the partial derivative $\partial_{X_a}$ acts locally on the vertex $a$ and effectively collapses all the internal lines that connect it to other vertices, generating source functions. The same phenomenon occurs for other vertices, which we can then combine together into the total differential. Continuing the process, we keep collapsing propagators until we are left with contributions containing only non-time-ordered propagators or a pure contact term. Interestingly, in doing this, we will see that multiple parent graphs can give rise to the same collapsed graphs when we contract them.  For example,  in the two-site case, the function $F_{\includegraphics{Figures/Subscripts/psic.pdf}}$ appeared in the differentials of both time orderings $\psi_{\includegraphics{Figures/Subscripts/psi+.pdf}}$ and $\psi_{\includegraphics{Figures/Subscripts/psi-.pdf}}$. This means that the same sources will appear in many places in the system of differential equations, and it is natural to consider the compatibilities between different ways of collapsing ordered graphs. Before discussing the general systematics, it is useful to first consider a slightly more complicated explicit example.

\paragraph{Three-site chain} 
An  instructive example is the case of the three-site chain.  The wavefunction now separates into $3^2=9$ parts:
\beq
\begin{aligned}
\psi \ =\ &\Big( \psi_{\includegraphics{Figures/Subscripts/psi++.pdf}}
\ +\
 \psi_{\includegraphics{Figures/Subscripts/psi+-.pdf}} \ +\
 \psi_{\includegraphics{Figures/Subscripts/psi--.pdf}} \ + \ 
 \psi_{\includegraphics{Figures/Subscripts/psi-+.pdf}} \Big) \\
 &\ - \
\Big( \psi_{\includegraphics{Figures/Subscripts/psi+0.pdf}} \ +\   \psi_{\includegraphics{Figures/Subscripts/psi-0.pdf}}\Big)
\ - \
\Big( \psi_{\includegraphics{Figures/Subscripts/psi0+.pdf}} \ +\   \psi_{\includegraphics{Figures/Subscripts/psi0-.pdf}}\Big) \ + \ \psi_{\includegraphics{Figures/Subscripts/psi00.pdf}} 
 \ ,
\end{aligned}
\label{equ:psi3}
\eeq
where the grouping into the $4+ 2 + 2+1$ terms is not accidental. 
The differential equations for these 9 functions are easy to derive (following the same procedure as above):
\begin{itemize}
\item The function with two non-time-ordered propagators closes on itself:
\be
{\rm d}  \psi_{\includegraphics{Figures/Subscripts/psi00.pdf}}   \, = \, \Big(
\alpha_1   \, {\rm d} \log(X_1^+)   \, +\, \alpha_2    \, {\rm d} \log(X_2^{++})  + \alpha_3    \, {\rm d} \log(X_3^+)   \Big) \, \psi_{\includegraphics{Figures/Subscripts/psi00.pdf}}\ , \label{equ:dpsi00}
\ee
where we defined the shorthand notation
\beq
\begin{aligned}
X_1^\pm &\equiv X_1 \pm Y_{12}\,, \quad
 X_2^{\pm \pm} \equiv X_2 \pm Y_{12} \pm Y_{23}\,,\quad
 X_3^\pm \equiv X_3 \pm Y_{23}\,,
 \end{aligned}
\eeq
with $Y_{ab}$ the internal energy flowing between vertices $a$ and $b$.
\item Next, we consider the functions with a single non-time-ordered propagator.  For example, the differentials of the functions $\psi_{\includegraphics{Figures/Subscripts/psi+0.pdf}}$ and $\psi_{\includegraphics{Figures/Subscripts/psi-0.pdf}}$ are
\begin{align}
&  \begin{aligned}
{\rm d}  \psi_{\includegraphics{Figures/Subscripts/psi+0.pdf}}  \ =\ &\Big( \alpha_1\,  \psi_{\includegraphics{Figures/Subscripts/psi+0.pdf}}  -  F_{\includegraphics{Figures/Subscripts/psic0.pdf}} \Big)\, {\rm d} \log(X_1^+)   \\
&+\, \Big( \alpha_2 \, \psi_{\includegraphics{Figures/Subscripts/psi+0.pdf}}  +  F_{\includegraphics{Figures/Subscripts/psic0.pdf}} \Big)\, {\rm d} \log(X_2^{-+})
\ +\ \alpha_3\,  \psi_{\includegraphics{Figures/Subscripts/psi+0.pdf}}  \ {\rm d} \log(X_3^{+}) \,,
\end{aligned}\label{equ:27} \\[10pt]
& \begin{aligned}
{\rm d}  \psi_{\includegraphics{Figures/Subscripts/psi-0.pdf}}  \ =\  &\Big( \alpha_1\,  \psi_{\includegraphics{Figures/Subscripts/psi-0.pdf}}  -  F_{\includegraphics{Figures/Subscripts/psic0.pdf}} \Big)\, {\rm d} \log(X_1^-)  \\
&+\, \Big( \alpha_2 \, \psi_{\includegraphics{Figures/Subscripts/psi-0.pdf}}  +  F_{\includegraphics{Figures/Subscripts/psic0.pdf}} \Big)\, {\rm d} \log(X_2^{++})
\ +\ \alpha_3\,  \psi_{\includegraphics{Figures/Subscripts/psi-0.pdf}}  \ {\rm d} \log(X_3^{+})\,,
\end{aligned}
\label{equ:28}
\end{align}
where the new source function $F_{\includegraphics{Figures/Subscripts/psic0.pdf}}$ arises from collapsing the propagator. Notice the similarity with~\eqref{equ:dpsi+} and~\eqref{equ:dpsi-} for the two-site chain. In particular, the {\it same} collapsed function $F_{\includegraphics{Figures/Subscripts/psic0.pdf}}$ appears in  the equations for both $\psi_{\includegraphics{Figures/Subscripts/psi+0.pdf}}$ and $\psi_{\includegraphics{Figures/Subscripts/psi-0.pdf}}$. 
As in~\eqref{equ:dpsic} for the two-site chain, the differential of $F_{\includegraphics{Figures/Subscripts/psic0.pdf}}$ closes on itself:
\beq\label{equ:dFc0}
{\rm d} F_{\includegraphics{Figures/Subscripts/psic0.pdf}} =  \Big(
(\alpha_1   \, +\, \alpha_2)    \, {\rm d} \log(X_{12})  + \alpha_3    \, {\rm d} \log(X_3^+)   \Big) \, F_{\includegraphics{Figures/Subscripts/psic0.pdf}}\ ,
\eeq
where $X_{12} \equiv X_{1}+X_{2}$.

Note that the differentials of the functions $\psi_{\includegraphics{Figures/Subscripts/psi+0.pdf}}$ and $\psi_{\includegraphics{Figures/Subscripts/psi-0.pdf}}$ only involve themselves and $F_{\includegraphics{Figures/Subscripts/psic0.pdf}}$, but not the functions $\psi_{\includegraphics{Figures/Subscripts/psi0+.pdf}}$ and  $\psi_{\includegraphics{Figures/Subscripts/psi0-.pdf}}$, which form their own closed sector along with $F_{\includegraphics{Figures/Subscripts/psi0c.pdf}}$, obeying similar equations. We therefore see that the contributions to the wavefunction with a single time-ordered propagator split into two independent sectors.

\item Finally, we look at the four functions without any non-time-ordered propagators. For example, the differential of the function $  \psi_{\includegraphics{Figures/Subscripts/psi++.pdf}} $ is
\beq
\begin{aligned}
{\rm d}  \psi_{\includegraphics{Figures/Subscripts/psi++.pdf}}  \ =\ &\Big( \alpha_1\,  \psi_{\includegraphics{Figures/Subscripts/psi++.pdf}}  -  F_{\includegraphics{Figures/Subscripts/psic+.pdf}} \Big)\, {\rm d} \log(X_1^+)  \\
&+\, \Big( \alpha_2 \, \psi_{\includegraphics{Figures/Subscripts/psi++.pdf}}  +  F_{\includegraphics{Figures/Subscripts/psic+.pdf}}  -  F_{\includegraphics{Figures/Subscripts/psi+c.pdf}} \Big)\, {\rm d} \log(X_2^{-+}) \\
&+\, \Big( \alpha_3\,  \psi_{\includegraphics{Figures/Subscripts/psi++.pdf}}  +  F_{\includegraphics{Figures/Subscripts/psi+c.pdf}} \Big)\, {\rm d} \log(X_3^-)\,,
\end{aligned} 
\label{equ:29}
\eeq
which has two source functions associated with the two ways of collapsing a propagator.
Similarly, the differential of the function $\psi_{\includegraphics{Figures/Subscripts/psi+-.pdf}}$ is
\beq
\begin{aligned}
{\rm d}  \psi_{\includegraphics{Figures/Subscripts/psi+-.pdf}}  \ =\ &\Big( \alpha_1\,  \psi_{\includegraphics{Figures/Subscripts/psi+-.pdf}}  -  F_{\includegraphics{Figures/Subscripts/psic-.pdf}} \Big)\, {\rm d} \log(X_1^+)  \\
&+\, \Big( \alpha_2 \, \psi_{\includegraphics{Figures/Subscripts/psi+-.pdf}}  +  F_{\includegraphics{Figures/Subscripts/psic-.pdf}}  +  F_{\includegraphics{Figures/Subscripts/psi+c.pdf}} \Big)\, {\rm d} \log(X_2^{--}) \\
&+\, \Big( \alpha_3\,  \psi_{\includegraphics{Figures/Subscripts/psi+-.pdf}}  -  F_{\includegraphics{Figures/Subscripts/psi+c.pdf}} \Big)\, {\rm d} \log(X_3^+)\,.
\end{aligned} 
\label{equ:30}
\eeq
The differentials of the other two time orderings are similar, and can be obtained by permutation.
Note that the same source function $F_{\includegraphics{Figures/Subscripts/psi+c.pdf}}$ appears in both~\eqref{equ:29} and~\eqref{equ:30}. The differential of this function is
\beq
{\rm d} F_{\includegraphics{Figures/Subscripts/psi+c.pdf}} \ =\  \Big( \alpha_1\,  F_{\includegraphics{Figures/Subscripts/psi+c.pdf}}  -  G_{\includegraphics{Figures/Subscripts/psicc.pdf}} \Big)\,  {\rm d} \log(X_1^+)  + \Big( (\alpha_2+\alpha_3)\,  F_{\includegraphics{Figures/Subscripts/psi+c.pdf}}  +  G_{\includegraphics{Figures/Subscripts/psicc.pdf}} \Big)\, {\rm d}\log(X_{23}^-)\,,
\label{equ:32}
\eeq
where $X_{23}^-\equiv X_2 + X_3 - Y_{12}$, and we have introduced a new source function $G_{\includegraphics{Figures/Subscripts/psicc.pdf}} $\,, with two collapsed propagators. The differentials of $F_{\includegraphics{Figures/Subscripts/psic+.pdf}}$ and $F_{\includegraphics{Figures/Subscripts/psic-.pdf}}$ can be obtained similarly.

Finally, the differential of the fully-collapsed function $G_{\includegraphics{Figures/Subscripts/psicc.pdf}}$ is
\beq
{\rm d} G_{\includegraphics{Figures/Subscripts/psicc.pdf}}  \ = \ (\alpha_1+\alpha_2 + \alpha_3) \,G_{\includegraphics{Figures/Subscripts/psicc.pdf}} \, {\rm d} \log(X_{123})\,,
\label{equ:33}
\eeq 
where $X_{123} \equiv X_1+X_2+X_3$.
\end{itemize}
It is clear from these examples---and from the more general arguments---that there is a pattern to this system of differential equations. In the following section, we will elucidate this pattern and give it a geometric interpretation.

\subsection{A Geometric Pattern}
\label{ssec:geometry}

The above equations have an interesting structure, relating the differentials of functions with time-ordered propagators to functions where these propagators have been collapsed. Since there are many possible paths to reach the same collapsed graph, we want to express the compatibilities between these different ways of collapsing internal lines. This will lead us to natural geometric structures associated with the differential equations that the wavefunction obeys.

\vskip4pt
To describe the pattern, it is useful to introduce a grading on the functions, corresponding to the  
number $r$ of non-time-ordered propagators. This grading separates independent sectors that aren't coupled in the differential system. Within each sector, functions are associated to directed graphs, which capture the possible time orderings. We have seen that contractions are natural from the time-integral perspective. The operation of contracting all edges between a given pair of vertices induces a partial ordering on the space of directed graphs. We will see that this poset can be geometrized by assigning the function/graphs to
components of simple geometrical shapes. 
We now illustrate this in a few examples.

\paragraph{Two-site chain} For the two-site chain, we have one function with $r=1$ and three functions with $r=0$, which can be arranged as
\beq
\raisebox{-1.25em}{ \begin{tikzpicture}[line width=0.75 pt, scale=0.75]
 \begin{scope}[xshift=-6cm]
\node at (0,0.5)  {\scalebox{1.5}{\includegraphics{Figures/Subscripts/psi0.pdf}}};
\draw[fill=Blue,Blue] (0,0) circle (5pt);
\end{scope}
\draw[fill=Blue,Blue] (-2,0) circle (5pt);
\draw[fill=Blue,Blue] (2,0) circle (5pt);
\draw[Blue,line width=2pt] (-2,0) -- (2,0);
\node at (-2,0.5)  {\scalebox{1.5}{\includegraphics{Figures/Subscripts/psi+.pdf}}};
\node at (2,0.5)  {\scalebox{1.5}{\includegraphics{Figures/Subscripts/psi-.pdf}}};
\node at (0,-0.4)  {\scalebox{1.5}{\includegraphics{Figures/Subscripts/psic.pdf}}};
\end{tikzpicture}}
\label{equ:2-geo}
\eeq
Here, the three $r=0$ functions have been assigned to the vertices and the edge of a line interval. 
The two disjoint geometrical objects in~\eqref{equ:2-geo} reflect the separation of the differential equation into independent $r=0$ and $r=1$ sectors.  Moreover, the assignment of functions on the interval captures geometrically how the ``edge function" $F_{\includegraphics{Figures/Subscripts/psic.pdf}}$, with the collapsed propagator, appears as a common source in the differential equation of the two ``vertex functions" $\psi_{\includegraphics{Figures/Subscripts/psi+.pdf}}$ and $\psi_{\includegraphics{Figures/Subscripts/psi-.pdf}}$.  

\paragraph{Three-site chain}
Similarly, for the three-site chain, we have one function with $r=2$, six functions with $r=1$, which further separate into two sectors, and nine functions with $r=0$:
\beq
\raisebox{-4.9em}{ \begin{tikzpicture}[line width=0.75 pt, scale=0.7]
 \begin{scope}[xshift=-5cm]
\node at (0,0.5)  {\scalebox{1.5}{\includegraphics{Figures/Subscripts/psi00.pdf}}};
\draw[fill=Blue,Blue] (0,0) circle (5pt);
\end{scope}
\begin{scope}[yshift=1.5cm]
\draw[fill=Blue,Blue] (-1.5,0) circle (5pt);
\draw[fill=Blue,Blue] (1.5,0) circle (5pt);
\draw[Blue,line width=2pt] (-1.5,0) -- (1.5,0);
\node at (-1.5,0.5)  {\scalebox{1.5}{\includegraphics{Figures/Subscripts/psi+0.pdf}}};
\node at (1.5,0.5)  {\scalebox{1.5}{\includegraphics{Figures/Subscripts/psi-0.pdf}}};
\node at (0,-0.4)  {\scalebox{1.5}{\includegraphics{Figures/Subscripts/psic0.pdf}}};
\end{scope}
\begin{scope}[yshift=-1.5cm]
\draw[fill=Blue,Blue] (-1.5,0) circle (5pt);
\draw[fill=Blue,Blue] (1.5,0) circle (5pt);
\draw[Blue,line width=2pt] (-1.5,0) -- (1.5,0);
\node at (-1.5,0.5)  {\scalebox{1.5}{\includegraphics{Figures/Subscripts/psi0+.pdf}}};
\node at (1.5,0.5)  {\scalebox{1.5}{\includegraphics{Figures/Subscripts/psi0-.pdf}}};
\node at (0,-0.4)  {\scalebox{1.5}{\includegraphics{Figures/Subscripts/psi0c.pdf}}};
\end{scope}
\begin{scope}[xshift=8cm]
\draw[fill=Blue,Blue] (-2,2) circle (5pt);
\draw[fill=Blue,Blue] (2,2) circle (5pt);
\draw[fill=Blue,Blue] (-2,-2) circle (5pt);
\draw[fill=Blue,Blue] (2,-2) circle (5pt);
\draw[fill=Blue, opacity=0.15] (-2,2) -- (2,2) -- (2,-2) -- (-2,-2) --cycle;
\draw[Blue,line width=2pt] (-2,2) -- (2,2) -- (2,-2) -- (-2,-2) --cycle;
\node at (0,0)  {\scalebox{1.5}{\includegraphics{Figures/Subscripts/psicc.pdf}}};
\node at (-2,2.5)  {\scalebox{1.5}{\includegraphics{Figures/Subscripts/psi++.pdf}}};
\node at (2,2.5)  {\scalebox{1.5}{\includegraphics{Figures/Subscripts/psi+-.pdf}}};
\node at (0,1.6)  {\scalebox{1.5}{\includegraphics{Figures/Subscripts/psi+c.pdf}}};
\node at (-2,-2.5)  {\scalebox{1.5}{\includegraphics{Figures/Subscripts/psi-+.pdf}}};
\node at (2,-2.5)  {\scalebox{1.5}{\includegraphics{Figures/Subscripts/psi--.pdf}}};
\node at (0,-1.6)  {\scalebox{1.5}{\includegraphics{Figures/Subscripts/psi-c.pdf}}};
\node at (-3,0)  {\scalebox{1.5}{\includegraphics{Figures/Subscripts/psic+.pdf}}};
\node at (3.,0)  {\scalebox{1.5}{\includegraphics{Figures/Subscripts/psic-.pdf}}};
\end{scope}
\end{tikzpicture}}
\label{equ:3-geo}
\eeq
Again, the disjoint geometrical objects (one point, two intervals, and one square) describe the separation of the differential equation into independent $r=0,1,2$ sectors. The edge functions~$F_i$ arise from collapsing the pair of time-ordered and anti-time-ordered propagators in the attached 
vertex functions~$\psi_i$. Similarly, the function associated to the face of the square, $G_{\includegraphics{Figures/Subscripts/psicc.pdf}}$,  is obtained by collapsing the remaining pairs of time-ordered propagators in the edge functions $F_i$.

\vskip 4pt 
Comparing~\eqref{equ:3-geo} with~\eqref{equ:2-geo} makes the recursive structure of the system manifest. The point and the two intervals in~\eqref{equ:3-geo}  arise when adding a non-time-ordered propagator to the two-site graph. Only the square is new and comes from adding time-ordered propagators. 

\newpage
\paragraph{Trees}
We are starting to see a pattern emerge. For each tree diagram with $e$ edges, there are ${}^e C_r \equiv {e\choose r}$ hypercubes of dimension $e-r$, to which the functions can be assigned.\footnote{A related observation was made in~\cite{De:2024zic}, who saw that the space of functions is naturally labeled by graph tubings corresponding to singularities of the wavefunction with coefficients that are scattering amplitudes. Since the presence of these singularities for a given subgraph is governed by the pattern of time-ordered propagators, the relation between the two organizations of the function basis is apparent. A nice observation of~\cite{De:2024zic} is that the functions can be decomposed into groups with elements counted by Pascal's triangle, independent of graph topology. This mirrors the topology independence of the collections of geometries associated to tree graphs.}

\vskip4pt
For example, 
in the case of the four-site graph (which has three edges), we have one function with $r=3$, nine functions with $r=2$ (which arrange themselves into three intervals), 27 functions with $r=1$ (arranged into three squares) and 27 functions with $r=0$ (forming a cube). The situation can be visualized as
\beq
\raisebox{-7.em}{ \begin{tikzpicture}[line width=2 pt, scale=0.7]
\begin{scope}[xshift=1cm]
\draw[fill=Blue,Blue] (-7,1) circle (5pt);
\node at (-7,1.5)  {\scalebox{1.4}{\includegraphics{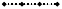}}};
\end{scope}
\draw[fill=Blue,Blue] (-1.75,4) circle (5pt);
\draw[fill=Blue,Blue] (1.75,4) circle (5pt);
\draw[Blue,line width=2pt] (-1.75,4) -- (1.75,4);
\node at (-3.5,4)  {\large $3 \times$};
\node at (-3.5,0)  {\large $3 \times$};
\node at (-1.75,4.5)  {\scalebox{1.4}{\includegraphics{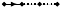}}};
\node at (1.75,4.5)  {\scalebox{1.4}{\includegraphics{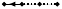}}};
\node at (0,3.6)  {\scalebox{1.4}{\includegraphics{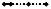}}};
%
%SQUARE
\draw[fill=Blue,Blue] (-1.75,1.75) circle (5pt);
\draw[fill=Blue,Blue] (1.75,1.75) circle (5pt);
\draw[fill=Blue,Blue] (-1.75,-1.75) circle (5pt);
\draw[fill=Blue,Blue] (1.75,-1.75) circle (5pt);
\draw[fill=Blue, opacity=0.15] (-1.75,1.75) -- (1.75,1.75) -- (1.75,-1.75) -- (-1.75,-1.75) --cycle;
\draw[Blue,line width=2pt] (-1.75,1.75) -- (1.75,1.75) -- (1.75,-1.75) -- (-1.75,-1.75) --cycle;
\node at (-1.75,2.25)  {\scalebox{1.4}{\includegraphics{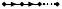}}};
\node at (1.75,2.25)  {\scalebox{1.4}{\includegraphics{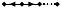}}};
\node at (0,1.35)  {\scalebox{1.4}{\includegraphics{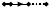}}};
\node at (-1.75,-2.25)  {\scalebox{1.4}{\includegraphics{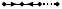}}};
\node at (1.75,-2.25)  {\scalebox{1.4}{\includegraphics{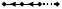}}};
\node at (0,-1.35)  {\scalebox{1.4}{\includegraphics{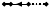}}};
\node at (0,0)  {\scalebox{1.4}{\includegraphics{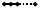}}};
%
%CUBE
\begin{scope}[xshift=5cm,yshift=-1.75cm]
	  \draw[color=Blue!50,dashed,line width=2pt] (1,1) -- (1,5);
  \draw[color=Blue!50,dashed] (5,1) -- (1,1);
    \draw[color=Blue!50,dashed] (0,0) -- (1,1);
	\draw[fill=Blue,opacity=0.15] (0,0) rectangle (4,4);
		\draw[color=Blue] (0,0) rectangle (4,4);
    \draw[fill=Blue,opacity=0.15] (4,0) -- (5,1) -- (5,5) -- (4,4) --cycle;
     \draw[color=Blue] (4,0) -- (5,1) -- (5,5) -- (4,4) --cycle;
       \draw[fill=Blue,opacity=0.15] (0,4) -- (4,4) -- (5,5) -- (1,5) --cycle;
          \draw[color=Blue] (0,4) -- (4,4) -- (5,5) -- (1,5) --cycle;           
 \draw[fill, color=Blue] (0,4) circle (5pt);	
 \draw[fill, color=Blue] (0,0) circle (5pt);
 \draw[fill, color=Blue] (4,4) circle (5pt);
 \draw[fill, color=Blue] (4,0) circle (5pt);
 \draw[fill, color=Blue] (1,5) circle (5pt);	
 \draw[fill, color=Blue] (1,1) circle (5pt);
 \draw[fill, color=Blue] (5,5) circle (5pt);
 \draw[fill, color=Blue] (5,1) circle (5pt);
\node at (5.5,0)  {\scalebox{1.4}{\includegraphics{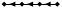}}};
\node at (2,-0.5)  {\scalebox{1.4}{\includegraphics{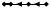}}};
\node at (6.5,1)  {\scalebox{1.4}{\includegraphics{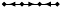}}};
\node at (5.9,0.5)  {\scalebox{1.4}{\includegraphics{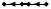}}};
\node at (2.5,0.5)  {\scalebox{1.4}{\includegraphics{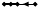}}};
\node at (2.25,2.5)  {\scalebox{1.4}{\includegraphics{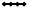}}};
\node at (6.5,5)  {\scalebox{1.4}{\includegraphics{Figures/Subscripts/psi-+-.pdf}}};
\node at (6.25,3)  {\scalebox{1.4}{\includegraphics{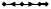}}};
\node at (2.5,6.25)  {\scalebox{1.4}{\includegraphics{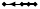}}};
\draw[-stealth,line width =1pt] (2.5,6.) -- (2.5,4.4);
 \end{scope}
\end{tikzpicture}}
\label{equ:4-geo}
\eeq
To avoid clutter, not all functions were drawn on the cube. Of the $64$ total functions appearing in the differential equations, only the $27$ functions associated to vertices sum together to give the wavefunction, 
while the functions assigned to the edges and faces come from collapsing time-ordered propagators.  Interestingly, this geometrical structure is independent of graph topology (at tree level), and both the star and chain have the same relationships between functions (though we have drawn the chain topology for concreteness). The difference in the result for the two actual wavefunctions is entirely driven by the different letters that appear in the differential equations.

\paragraph{Loops}
We can find similar patterns in the differential equations satisfied by loops~\cite{Baumann:2024mvm,Hang:2024xas}. The essential novelty for loop diagrams is that we have to collapse all propagators that connect two vertices at the same time, so the compatibilities of these patterns of collapsing propagators are related to new geometric shapes beyond hypercubes, which interestingly depend on the specific topology considered.

\begin{itemize}
\item {\it Bubble:} \
The simplest one-loop example is the two-site bubble:
 \begin{equation}
\raisebox{-1.35em}{\begin{tikzpicture}[line width=1. pt, scale=1.8]
\draw[black] (0.35,0) circle (.35cm);
\draw[fill=black,black] (0,0) circle (.03cm);
\draw[fill=black,black] (0.7,0) circle (.03cm);
\end{tikzpicture} } \end{equation}
 This diagram has two edges and so we might naively expect it to have 16 basis elements, just like for the three-site chain. However, not all time orderings are permitted for this diagram:  If we fix one time ordering using the top line, say $\eta_1 >\eta_2$, then we cannot enforce $\eta_2 <\eta_1$ on the bottom line as this would lead to a contradiction. The following two ``cyclic time orderings" are therefore forbidden:
  \begin{equation}
\includegraphics[valign=c]{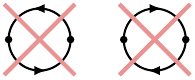}
\label{equ:forbidden}
\end{equation}
Hence, only 7 (not $3^2=9$) terms contribute to the wavefunction:
 \begin{equation}
\includegraphics[valign=c]{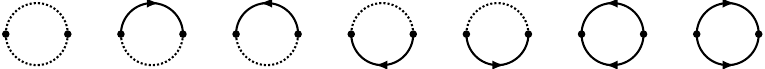}
\label{equ:bubble}
\end{equation}
In the language of graph theory, this means that we only consider {\it directed acyclic graphs}.

When we take derivatives of these contributions, we again collapse pairs of time-ordered propagators. For the last two terms in~\eqref{equ:bubble}, with two ordered propagators, we need to consider that collapsing a single propagator also effectively collapses the other one. This leads to the following geometric arrangement relating the various functions:
\beq
\includegraphics[valign=c]{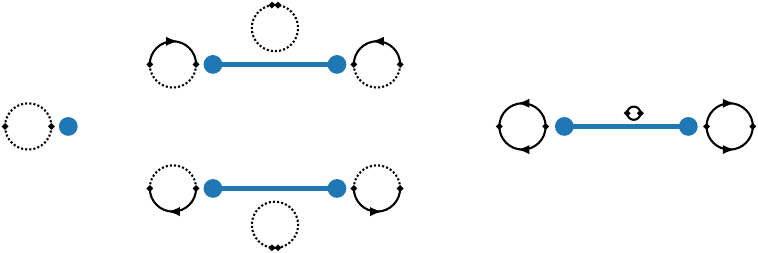}
\label{equ:bubble-geo}
\eeq
The isolated point and the first two intervals are the same structures as  in~\eqref{equ:3-geo} for the three-site chain. 
At the level of the integrand, these contributions are the same as tree-level diagrams in which the energy of each vertex is shifted  by the sum of the energies associated to the non-time-ordered lines. Hence, the differential equations for these contributions can be deduced simply from~\eqref{equ:dpsi0}--\eqref{equ:dpsic}. 

If it were possible to contract the two internal lines separately, the completely time-ordered contributions would naturally be associated to the vertices of a square as in~\eqref{equ:3-geo}. We can understand the reduction to a line interval by taking into account that the two vanishing time orderings in (\ref{equ:forbidden}). This removes two vertices and the attached edges from the square, and we are left with an interval:
\beq
\includegraphics[valign=c]{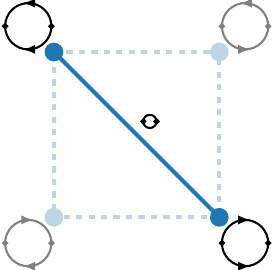}
\eeq
We can alternatively view this as a projection of the square into the line defined by connecting the two nonzero time-ordered functions.

In summary, the differential equations for the one-loop bubble separates into three groups of three functions that close among themselves plus a totally un-time-ordered piece. This is exactly as expected from the tree theorem of~\cite{ManuelThesis,AguiSalcedo:2023nds}, and reproduces the differential equations found in~\cite{Baumann:2024mvm,Hang:2024xas} after an appropriate transformation of the basis.

\item {\it Triangle:}\
As another example, we consider the one-loop triangle graph:
\begin{equation}
\begin{aligned}
	\begin{tikzpicture}[line width=1. pt, scale=1.8]
		\timeloop{0}{0}{.35cm}{3}{.03cm}{3,3,3}{1,1,1}{1}	
	\end{tikzpicture}
\end{aligned}
\end{equation}
Once again, there are two (cyclic) time orderings that vanish, giving us $3^3-2=25$ contributions to the wavefunction. These come in four different types graded by the number of disconnected lines in the loops:
\begin{equation}\label{equ:Trigram}
\includegraphics[valign=c]{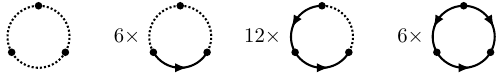}
\end{equation}
Derivatives again collapse time-ordered propagators to produce source functions. For the functions containing at least one disconnected edge this produces the same geometric arrangement as for the equivalent functions in the four-site case; cf.~\eqref{equ:4-geo}. The only novelty is the fully-connected contribution which only has six basis functions rather than eight. Moreover, every diagram will have exactly one edge that, if collapsed, leaves the other two edges with a forbidden time ordering.  
Each parent function is therefore only connected to two source functions rather than three. Similarly, after collapsing one of these edges the remaining two must also be collapsed simultaneously. The functions in the fully-connected sector thus combine into a hexagon. The complete geometric arrangement of the basis functions  is 
\beq
\includegraphics[valign=c]{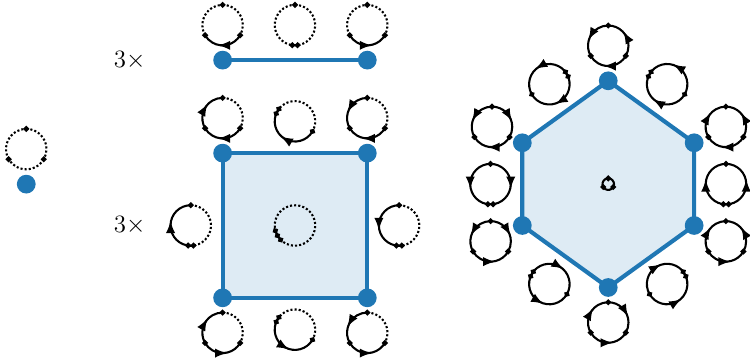}
\label{equ:triangle-geo}
\eeq
We can also think of the hexagon as arising from the cube in the  tree-level four-site example after removing the vertices  corresponding to forbidden time orderings (as well as the attached edges and faces):
\beq\label{equ:triangle-cube}
\includegraphics[valign=c]{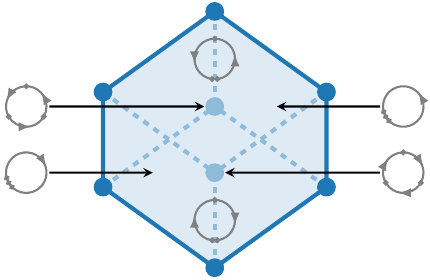}
\eeq
Note that the removal of these two vertices projects the cube along the diagonal joining those vertices together (or, equivalently, onto the plane in which all the nonzero vertices lie), which reduces the dimension of the final shape. This is the generic behavior for an $n$-cycle---and indeed for all other loop diagrams (although not all projections will be along the same diagonal). The class of polytopes that can be produced from the projections of hypercubes are known as zonotopes~\cite{postnikov2009permutohedra} and any differential system will arrange itself into exactly $2^e$ of them. The precise polytopes appearing will depend on the graph topology.

\item {\it Higher loops:} One can easily extend the treatment to the integrands of higher-order loops. As before, any basis element with a cyclic time ordering will vanish, reducing the size of the geometric objects associated to the basis functions.

As an example, consider two-site, $n$-loop ``banana" diagrams (which were recently studied in~\cite{Westerdijk:2025ywh}). Such graphs have $n+1$ internal edges. As for the one-loop bubble, if we collapse a single time-ordered propagator, we must collapse them all. This implies that the geometry of interest is a collection of line intervals and one isolated point. We can still grade the decoupled sectors by their number of non-time-ordered propagators, $r$. In each sector, there are ${e\choose r}$ different ways of picking these non-time-ordered propagators, and each leads to a disjoint line interval. (The exception is the totally non-time-ordered graph, which is a single function, corresponding to a point.) Summing all sectors, we conclude that we need $2^n-1$ line segments and a single vertex resulting in a total of $1+3(2^{n+1}-1)$ master integrals. 
\end{itemize}
Now that we have understood how to organize the space of functions geometrically, we would like to develop an efficient procedure to read off the differential equations themselves from these geometric structures. We will do this in the next section.

%%%%%%%%%%%%%%%%%%%%%%%%%%%%%%%%%%%%%%%%%%%%%%%%%%%%%%%%%
\newpage
\section{Boundary Kinematics}
\label{sec:boundary}

We have seen that the differential equations satisfied by the time-ordered components of the wavefunction are naturally related to geometric objects that encode different ways to collapse edges of the underlying Feynman graphs.
In this section, we will show that the same physics can also be described purely in terms of the kinematics on the boundary.  As in~\cite{Arkani-Hamed:2023kig}, we introduce graph tubings to represent both the letters appearing in the equations and the basis functions themselves.  The critical difference is that the objects involved will represent a different set of basis functions. We will see that this alternative choice of basis greatly simplifies the structure of the differential equations and the associated kinematic flow rules. 

\subsection{Graph Tubings}
\label{ssec:tubings}

We begin by describing the representation of letters and functions in terms of the graph tubings introduced in~\cite{Arkani-Hamed:2023kig}.  Given a graph, a {\it tube} is a connected subgraph. Tubes have a natural pictorial representation---we circle the vertices and edges of the relevant subgraph.
There is also a natural notion of compatibility between tubes. We say that two tubes are compatible if their intersection is empty (pictorially, if the circles do not cross).\footnote{Note that our definition of tube, and our notion of compatibility differ from that in the mathematics literature~\cite{CARR20062155}. Typically, tubes are required to be {\it proper} subgraphs, but it will be desirable for us to consider the graph itself as part of the construction. Further, in the math literature adjacent tubes are typically considered to not be compatible, but we will want to allow such configurations to occur. In fact, such adjacent tubes will play a critical role in the kinematic flow.}
We additionally introduce the notion of a {\it marked graph}. Given a graph, we generate its marked counterpart by decorating all of its internal lines with a cross. With respect to tubings, the cross behaves like an ordinary vertex {\it except} it cannot be part of a subgraph itself without any ordinary vertices (i.e.~we cannot encircle only a cross without any vertices within a tube).

\vskip4pt
We now describe the representation of letters and functions in terms of marked graphs. Letters correspond to single tubes, while functions will be (suitably defined) complete tubings.

\paragraph{Letters} 
We first describe how to label letters.
Starting from the original Feynman graph, we obtain its associated marked graph by adding a cross to every internal line. 
The letters are then represented by single tubes (single circlings of the vertices and crosses) of these marked graphs. Recall that each tube must contain at least one vertex. To each tubing, we then assign a letter given by the sum of vertex energies enclosed by the tube and the energies of the internal lines piercing the tube. For tubes that intersect an internal line and enclose the corresponding cross, we flip the sign of the internal energy.

\vskip 4pt
In order to make this more explicit, we enumerate the letters of the two-site chain\footnote{Strictly speaking, the tubings in~\eqref{equ:Letters-2Site}
 represent dlog-forms with the letters appearing as their arguments rather than the letters themselves.  For simplicity, we will continue to use this slightly imprecise language.} 
\beq
\begin{aligned}
	&\includegraphics[scale=0.9,valign=c]{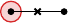}  \ \equiv\ \ud\log(X_1+Y)\,,  &\quad&\includegraphics[scale=0.9,valign=c]{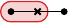} \ \equiv\ \ud\log(X_1-Y)\,,\\
	&\includegraphics[scale=0.9,valign=c]{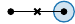}  \ \equiv\ \ud\log(X_2+Y)\,, &&\includegraphics[scale=0.9,valign=c]{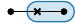} \ \equiv\ \ud\log(X_2-Y)\,,\\
	&\includegraphics[scale=0.9,valign=c]{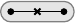}  \ \equiv\ \ud\log(X_1+X_2)\,.
\end{aligned}\label{equ:Letters-2Site}
\eeq
The letters in the left column correspond to the known energy singularities of the graph, while those on the right come from a sign flip of the internal energy. 
Their appearance is a consequence of the fact that the FRW wavefunction can be viewed as an integral of the flat-space wavefunction, whose singularities are those in the left column~\cite{Arkani-Hamed:2023kig}. Upon integration, the new singularities shown in the right column can appear.

\vskip4pt
Similarly, the letters for the three-site chain are
\beq
\begin{aligned}
	&\includegraphics[scale=0.9,valign=c]{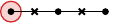} \ \equiv\ \ud\log(X_1+Y_{12}) \,, &\quad& \includegraphics[scale=0.9,valign=c]{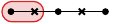}\ \equiv\ \ud\log(X_1-Y_{12})\,,\\
	&\includegraphics[scale=0.9,valign=c]{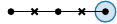} \ \equiv\ \ud\log(X_3+Y_{23})\,, &&\includegraphics[scale=0.9,valign=c]{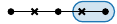}\ \equiv\ \ud\log(X_3-Y_{23})\,,\\
	&\includegraphics[scale=0.9,valign=c]{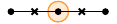} \ \equiv\ \ud\log(X_2+Y_{12}+Y_{23}) \,,&& \includegraphics[scale=0.9,valign=c]{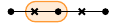}\ \equiv\ \ud\log(X_2-Y_{12}+Y_{23})\,,\\
	& && \includegraphics[scale=0.9,valign=c]{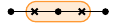}\ \equiv\ \ud\log(X_2-Y_{12}-Y_{23})\,,\\
	& \includegraphics[scale=0.9,valign=c]{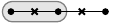}\ \equiv\ \ud\log(X_1+X_2+Y_{23})\,,&& \includegraphics[scale=0.9,valign=c]{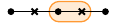}\ \equiv\ \ud\log(X_2+Y_{12}-Y_{23})\,,\\
	&\includegraphics[scale=0.9,valign=c]{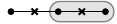}\ \equiv\ \ud\log(X_2+X_3+Y_{12})\,, && \includegraphics[scale=0.9,valign=c]{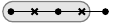}\ \equiv\ \ud\log(X_1+X_2-Y_{23})\,,\\
	& \includegraphics[scale=0.9,valign=c]{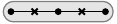}\ \equiv\ \ud\log(X_1+X_2+X_3)\,, \quad && \includegraphics[scale=0.9,valign=c]{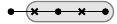}\ \equiv\ \ud\log(X_2+X_3-Y_{12})\,.
\end{aligned}\label{equ:3Site-Letters}
\eeq
Again, we have separated the contributions into the known energy singularities of the graph on the left and the new singularities with flipped sign of some internal energy on the right.

\paragraph{Functions} The functions appearing in the differential system can also be represented by tubings. We first define a {\it complete tubing} of a marked graph following~\cite{Arkani-Hamed:2023kig} as a collection of compatible tubes where each vertex is a member of exactly one tube (but crosses need not be encircled). The functions in the differential system are then in one-to-one correspondence with the complete tubings of the associated marked graph.
\begin{itemize}
\item {\it Two-site chain}\\[4pt]
Our first example is the two-site chain.
There are four different ways to create a complete tubing of the marked graph:
\beq
\includegraphics[scale=0.9,valign=c]{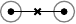}  \qquad
\includegraphics[scale=0.9,valign=c]{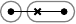} \qquad
\includegraphics[scale=0.9,valign=c]{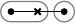}  \qquad 
\includegraphics[scale=0.9,valign=c]{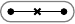} 
\eeq
Note that the tubings of the marked graph induce a directionality that naturally matches the two time-orderings of the bulk propagator:
\beq
\begin{aligned}
&\includegraphics[scale=0.9,valign=c]{Figures/Tubings/Functions/psi+.pdf} 
 \ \leftrightarrow\  \includegraphics[scale=1.4,valign=c]{Figures/Subscripts/psi+.pdf} \\
&\includegraphics[scale=0.9,valign=c]{Figures/Tubings/Functions/psi-.pdf}
 \ \leftrightarrow\  \includegraphics[scale=1.4,valign=c]{Figures/Subscripts/psi-.pdf}  
\end{aligned}
\label{equ:time-ordering}
\eeq
The tubing with the exposed cross is in correspondence to the non-time-ordered propagator:
\beq
\includegraphics[scale=0.9,valign=c]{Figures/Tubings/Functions/psi0.pdf} 
 \ \leftrightarrow\  \includegraphics[scale=1.4,valign=c]{Figures/Subscripts/psi0.pdf}  
\label{equ:non-time-ordering}
\eeq
Finally, the tubing encircling both vertices corresponds to the collapsed propagator:
\beq
\includegraphics[scale=0.9,valign=c]{Figures/Tubings/Functions/psic.pdf} 
 \ \leftrightarrow\  \includegraphics[scale=1.4,valign=c]{Figures/Subscripts/psic.pdf}  \hspace{0.4cm}\phantom{x}
\label{equ:collapsed}
\eeq
As  in Section~\ref{ssec:geometry}, it will be useful to represent the relations between the tubings by the following geometrical objects: 
\beq
\raisebox{-1.2em}
 {\begin{tikzpicture}[line width=0.75 pt, scale=0.65]
 \begin{scope}[xshift=-6cm]
\node at (0,0.5)  {\includegraphics[scale=0.8,valign=c]{Figures/Tubings/Functions/psi0.pdf}};
\draw[fill=Red,Red] (0,0) circle (5pt);
\end{scope}
\draw[fill=Orange,Orange] (-2,0) circle (5pt);
\draw[fill=Orange,Orange] (2,0) circle (5pt);
\draw[Orange,line width=2pt] (-2,0) -- (2,0);
\node at (-2,0.5)  {\includegraphics[scale=0.8,valign=c]{Figures/Tubings/Functions/psi+.pdf}};
\node at (2,0.5)  {\includegraphics[scale=0.8,valign=c]{Figures/Tubings/Functions/psi-.pdf}};
\node at (0,-0.4)  {\includegraphics[scale=0.8,valign=c]{Figures/Tubings/Functions/psic.pdf}};
\end{tikzpicture}}
\label{equ:2site-geometry}
\eeq
The function $\includegraphics[scale=0.7,valign=c]{Figures/Tubings/Functions/psi0.pdf}$ satisfies its own differential equation; i.e.~it is decoupled from the other functions.
The function $\includegraphics[scale=0.7,valign=c]{Figures/Tubings/Functions/psic.pdf}$ (assigned to the edge of the interval) appears as a source function in the differential equations
of the functions $\includegraphics[scale=0.7,valign=c]{Figures/Tubings/Functions/psi+.pdf}$ and $\includegraphics[scale=0.7,valign=c]{Figures/Tubings/Functions/psi-.pdf}$ (assigned to the attached vertices). 
We can think of the edge function  as arising from the merger of the tubes of the connected vertex functions.

To each complete tubing, we assign a specific basis function. These basis functions can be understood directly from the time-integral representation using the identifications~\eqref{equ:time-ordering}. However, it is illuminating to express them in terms of energy integrals defined purely on the boundary of the spacetime~\cite{Arkani-Hamed:2017fdk,Arkani-Hamed:2023kig}. 
To map the time integrals of Section~\ref{sec:BulkTime} to these energy integrals, we use the identity
\begin{align}
    \int \ud\eta\, \frac{e^{iX\eta}}{(-\eta)^{1+\alpha}} =\frac{e^{\frac{i\pi \alpha}{2}}}{\Gamma(1+\alpha)}\int_0^\infty \textrm{d}\omega \,\omega^\alpha \frac{1}{\omega+X}\,.
\end{align}
Performing the time integrals~\eqref{equ:psiT}--\eqref{equ:psiD} by using this substitution, we get
\begin{align}
\psi_{\includegraphics[scale=0.6]{Figures/Tubings/Functions/psi0.pdf}}  & = N \int {\rm d} \omega_1\hs {\rm d} \omega_2 \,\omega_1^{\alpha_1} \omega_2^{\alpha_2}\, \frac{1}{(X_1+\omega_1+Y) (X_2+\omega_2+Y)}\,, 
\label{equ:4-psi0} \\
\psi_{\includegraphics[scale=0.6]{Figures/Tubings/Functions/psi+.pdf}}  & = N \int {\rm d} \omega_1\hs {\rm d} \omega_2  \,\omega_1^{\alpha_1} \omega_2^{\alpha_2} \, \frac{1}{(X_1+X_2+\omega_1+\omega_2)(X_1+\omega_1+Y) }\,, \label{equ:4-psi+}\\
\psi_{\includegraphics[scale=0.6]{Figures/Tubings/Functions/psi-.pdf}} &= N \int {\rm d} \omega_1\hs {\rm d} \omega_2  \,\omega_1^{\alpha_1} \omega_2^{\alpha_2}\, \frac{1}{(X_1+X_2+\omega_1+\omega_2)(X_2+\omega_2+Y) }\,. \label{equ:4-psi-}
\end{align}
Note that the integrated energies, $\omega_i$, always comes in combination with the vertex energies as $X_i+\omega_i$. Moreover, the rational part of the integrands in the functions $\psi_i$ is the flat-space wavefunction (separated into three pieces). Integrated against the twist factor $\omega_i^{\alpha_i}$, this transforms the flat-space results to the results in a power-law cosmology.
The wavefunction itself is
\beq
\begin{aligned}
\psi &\ =\ \psi_{\includegraphics[scale=0.6]{Figures/Tubings/Functions/psi+.pdf}}
\ +\
\psi_{\includegraphics[scale=0.6]{Figures/Tubings/Functions/psi-.pdf}}
\ - \
\psi_{\includegraphics[scale=0.6]{Figures/Tubings/Functions/psi0.pdf}} \\[4pt]
 &\ = \ N \int {\rm d} \omega_1\hs {\rm d} \omega_2  \,\omega_1^{\alpha_1} \omega_2^{\alpha_2}\, \frac{2Y}{(X_1+X_2+\omega_1+\omega_2)(X_1+\omega_1+Y)(X_2+\omega_2+Y)} \, .
\end{aligned}
\label{equ:psi-total}
\eeq
Similar manipulations for the integral (\ref{equ:psiD}) lead to 
\begin{align}
	F_{\includegraphics[scale=0.6]{Figures/Tubings/Functions/psic.pdf}} &=\int \frac{\textrm{d} \eta_1 \textrm{d}\eta_2}{(-\eta_1)^{1+\alpha_1} (-\eta_2)^{1+\alpha_2}}e^{iX_1\eta_1}e^{iX_2\eta_2}\, \eta_1 \delta(\eta_1-\eta_2) \nonumber \\
	&=i\alpha_1 N \int  {\rm d} \omega_1\hs {\rm d} \omega_2  \,\omega_1^{\alpha_1-1} \omega_2^{\alpha_2} \,\frac{-i}{X_1+X_2+\omega_1+\omega_2} \nonumber \\
	&=N \int {\rm d} \omega_1\hs {\rm d} \omega_2  \,\omega_1^{\alpha_1} \omega_2^{\alpha_2} \,\frac{1}{(X_1+X_2+\omega_1+\omega_2)^2}\,, \label{equ:4-F}
\end{align}
where the final equality  follows from integration by parts in $\omega_1$.  Note that the pole in the integrand is second order, which is unlike the usual poles of the flat-space wavefunction (for polynomial interactions), which are first order.

Taking derivatives with respect to the variables $X_1$ and $X_2$, it is straightforward to derive the differential equations satisfied by these energy integrals. The result can be given the graphical representation
\begin{align}
{\rm d}  \psi_{\includegraphics[scale=0.6]{Figures/Tubings/Functions/psi0.pdf}}   &\, = \, \Big(
\alpha_1   \, \includegraphics[scale=0.9,valign=c]{Figures/Tubings/two/Lap.pdf}   \, +\, \alpha_2     \includegraphics[scale=0.9,valign=c]{Figures/Tubings/two/Lbp.pdf}  \Big) \,  \psi_{\includegraphics[scale=0.6]{Figures/Tubings/Functions/psi0.pdf}} \ ,  \label{equ:4-3}\\
{\rm d}\psi_{\includegraphics[scale=0.6]{Figures/Tubings/Functions/psi+.pdf}}  &\, = \, \Big(
\alpha_1   \, \includegraphics[scale=0.9,valign=c]{Figures/Tubings/two/Lap.pdf}   \, +\, \alpha_2     \includegraphics[scale=0.9,valign=c]{Figures/Tubings/two/Lbm.pdf}  \Big) \,  \psi_{\includegraphics[scale=0.6]{Figures/Tubings/Functions/psi+.pdf}} 
  \, +\, 
  \Big(\includegraphics[scale=0.9,valign=c]{Figures/Tubings/two/Lbm.pdf}   \, -\,     \includegraphics[scale=0.9,valign=c]{Figures/Tubings/two/Lap.pdf}  \Big)  \, F_{\includegraphics[scale=0.6]{Figures/Tubings/Functions/psic.pdf}}   \,, \label{equ:4-4}\\
{\rm d}\psi_{\includegraphics[scale=0.6]{Figures/Tubings/Functions/psi-.pdf}} &\, = \, \Big(
\alpha_1   \, \includegraphics[scale=0.9,valign=c]{Figures/Tubings/two/Lam.pdf}   \, +\, \alpha_2     \includegraphics[scale=0.9,valign=c]{Figures/Tubings/two/Lbp.pdf}  \Big) \,  \psi_{\includegraphics[scale=0.6]{Figures/Tubings/Functions/psi+.pdf}} 
  \, +\, 
  \Big(
 \includegraphics[scale=0.9,valign=c]{Figures/Tubings/two/Lam.pdf}   \, -\,     \includegraphics[scale=0.9,valign=c]{Figures/Tubings/two/Lbp.pdf}  \Big)  \, F_{\includegraphics[scale=0.6]{Figures/Tubings/Functions/psic.pdf}}  \,, \label{equ:4-5}\\[2pt]
{\rm d} 
F_{\includegraphics[scale=0.6]{Figures/Tubings/Functions/psic.pdf}}  &\, = \, (\alpha_1 + \alpha_2)\, 
 \includegraphics[scale=0.9,valign=c]{Figures/Tubings/two/Lab.pdf}\, F_{\includegraphics[scale=0.6]{Figures/Tubings/Functions/psic.pdf}} \,.
\label{equ:4-6}
\end{align}
Of course, 
these are equivalent to equations~\eqref{equ:dpsi0}--\eqref{equ:dpsic} in Section~\ref{ssec:DE}, if we apply the dictionary~\eqref{equ:time-ordering}--\eqref{equ:collapsed}.
Defining the vector 
\be
	\vec I = ( \red{\psi_{\includegraphics[scale=0.6]{Figures/Tubings/Functions/psi0.pdf}}},\,\orange{\psi_{\includegraphics[scale=0.6]{Figures/Tubings/Functions/psi+.pdf}} },\,\orange{\psi_{\includegraphics[scale=0.6,reflect]{Figures/Tubings/Functions/psi+.pdf}}},\,\orange{F_{\includegraphics[scale=0.6]{Figures/Tubings/Functions/psic.pdf}}})\,,
\ee
the differential equation can be written as a matrix equation $\ud \vec{I} =  A \,\vec{I}$, where
the nonzero elements of the connection matrix $A$ can be visualized as
 \beq
A = \left[\includegraphics[valign=c,trim={0 5pt 0 5pt 0}]{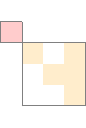}\right].
\label{equ:2site-Amatrix}
\eeq
Notice that this matrix separates into two blocks (one of which involves a single function) that only mix within themselves when taking a differential.
This pattern in~\eqref{equ:2site-Amatrix} reflects the fact that the basis functions separate into two geometric objects in~\eqref{equ:2site-geometry}. The matrix is upper-triangular, which makes the differential equation particularly easy to solve (see Section~\ref{ssec:BC}). These features will be present in all examples.

%\newpage
\item {\it Three-site chain}\\[4pt] 
For the three-site chain, the set of complete tubings and associated basis functions are 
\beq
\begin{aligned}
&\includegraphics[scale=0.9,valign=c]{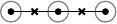}  \ \leftrightarrow\   \includegraphics[scale=1.4,valign=c]{Figures/Subscripts/psi00.pdf}\ &\quad &  
\includegraphics[scale=0.9,valign=c,reflect]{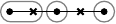}
 \ \leftrightarrow\   \includegraphics[scale=1.4,valign=c]{Figures/Subscripts/psi0+.pdf} &\quad &
\includegraphics[scale=0.9,valign=c]{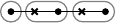}  \ \leftrightarrow\   \includegraphics[scale=1.4,valign=c]{Figures/Subscripts/psi++.pdf} \\
&&\quad & 
\includegraphics[scale=0.9,valign=c,reflect]{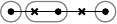}
  \ \leftrightarrow\   \includegraphics[scale=1.4,valign=c]{Figures/Subscripts/psi0-.pdf} &\quad &
\includegraphics[scale=0.9,valign=c,reflect]{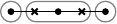}  \ \leftrightarrow\   \includegraphics[scale=1.4,valign=c]{Figures/Subscripts/psi+-.pdf} \\
&&\quad &  \includegraphics[scale=0.9,valign=c,reflect]{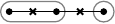} \ \leftrightarrow\   \includegraphics[scale=1.4,valign=c,reflect]{Figures/Subscripts/psic0.pdf} &\qquad &
\includegraphics[scale=0.9,valign=c,reflect]{Figures/Tubings/Functions/psi++.pdf}  \ \leftrightarrow\   \includegraphics[scale=1.4,valign=c]{Figures/Subscripts/psi--.pdf} \\
&&&&\quad &
\includegraphics[scale=0.9,valign=c]{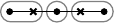}  \ \leftrightarrow\   \includegraphics[scale=1.4,valign=c]{Figures/Subscripts/psi-+.pdf} \\
&&\quad &  \includegraphics[scale=0.9,valign=c]{Figures/Tubings/Functions/psi+0.pdf}  \ \leftrightarrow\   \includegraphics[scale=1.4,valign=c,reflect]{Figures/Subscripts/psi0-.pdf} &\quad &
\includegraphics[scale=0.9,valign=c]{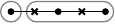}  \ \leftrightarrow\   \includegraphics[scale=1.4,valign=c,reflect]{Figures/Subscripts/psic-.pdf} \\
&&\quad &  \includegraphics[scale=0.9,valign=c]{Figures/Tubings/Functions/psi-0.pdf}  \ \leftrightarrow\   \includegraphics[scale=1.4,valign=c,reflect]{Figures/Subscripts/psi0+.pdf}&\quad &
\includegraphics[scale=0.9,valign=c,reflect]{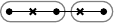}  \ \leftrightarrow\   \includegraphics[scale=1.4,valign=c,reflect]{Figures/Subscripts/psic+.pdf} \\
&&\quad &  \includegraphics[scale=0.9,valign=c]{Figures/Tubings/Functions/psic0.pdf}  \ \leftrightarrow\   \includegraphics[scale=1.4,valign=c]{Figures/Subscripts/psic0.pdf}  &\quad &
\includegraphics[scale=0.9,valign=c]{Figures/Tubings/Functions/psic+.pdf} \ \leftrightarrow\   \includegraphics[scale=1.4,valign=c]{Figures/Subscripts/psic+.pdf} \\
&&\quad &   &\qquad &
\includegraphics[scale=0.9,valign=c,reflect]{Figures/Tubings/Functions/psi+c.pdf}   \ \leftrightarrow\   \includegraphics[scale=1.4,valign=c]{Figures/Subscripts/psic-.pdf} \\
&&\quad &   &\qquad &
\includegraphics[scale=0.9,valign=c]{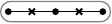}  \ \, \leftrightarrow\   \includegraphics[scale=1.4,valign=c]{Figures/Subscripts/psicc.pdf} 
\end{aligned}
\label{equ:Functions-2Site}
\eeq
The functions have been graded by the number of unenclosed crosses. This is in one-to-one correspondence with the number of non-time-ordered propagators. 

 It is again useful to arrange the functions geometrically to summarize how they appear in the differential equations: 
\beq
 \raisebox{-4.6em}{
 \begin{tikzpicture}[line width=0.75 pt, scale=0.65]
 \begin{scope}[xshift=-5.5cm]
\node at (0,0.5)  {\includegraphics[scale=0.8,valign=c]{Figures/Tubings/Functions/psi00.pdf}};
\draw[fill=Red,Red] (0,0) circle (5pt);
\end{scope}
\begin{scope}[yshift=1.5cm]
\draw[fill=Orange,Orange] (-2,0) circle (5pt);
\draw[fill=Orange,Orange] (2,0) circle (5pt);
\draw[Orange,line width=2pt] (-2,0) -- (2,0);
\node at (-2,0.5)  {\includegraphics[scale=0.8,valign=c]{Figures/Tubings/Functions/psi+0.pdf}};
\node at (2,0.5)  {\includegraphics[scale=0.8,valign=c]{Figures/Tubings/Functions/psi-0.pdf}};
\node at (0,-0.4)  {\includegraphics[scale=0.8,valign=c]{Figures/Tubings/Functions/psic0.pdf}};
\end{scope}
\begin{scope}[yshift=-1.5cm]
\draw[fill=Orange,Orange] (-2,0) circle (5pt);
\draw[fill=Orange,Orange] (2,0) circle (5pt);
\draw[Orange,line width=2pt] (-2,0) -- (2,0);
\node at (-2,0.5)  {\includegraphics[scale=0.8,valign=c,reflect]{Figures/Tubings/Functions/psi-0.pdf}};
\node at (2,0.5)  {\includegraphics[scale=0.8,valign=c,reflect]{Figures/Tubings/Functions/psi+0.pdf}};
\node at (0,-0.4)  {\includegraphics[scale=0.8,valign=c,reflect]{Figures/Tubings/Functions/psic0.pdf}};
\end{scope}
\begin{scope}[xshift=9cm]
\draw[fill=Green,Green] (-2,2) circle (5pt);
\draw[fill=Green,Green] (2,2) circle (5pt);
\draw[fill=Green,Green] (-2,-2) circle (5pt);
\draw[fill=Green,Green] (2,-2) circle (5pt);
\draw[fill=Green, opacity=0.15] (-2,2) -- (2,2) -- (2,-2) -- (-2,-2) --cycle;
\draw[Green,line width=2pt] (-2,2) -- (2,2) -- (2,-2) -- (-2,-2) --cycle;
\node at (0,0)  {\includegraphics[scale=0.8,valign=c,reflect]{Figures/Tubings/Functions/psicc.pdf}};
\node at (-2,2.5)  {\includegraphics[scale=0.8,valign=c]{Figures/Tubings/Functions/psi++.pdf}};
\node at (2,2.5)  {\includegraphics[scale=0.8,valign=c]{Figures/Tubings/Functions/psi-+.pdf}};
\node at (0,1.6)  {\includegraphics[scale=0.8,valign=c]{Figures/Tubings/Functions/psi+c.pdf}};
\node at (-2,-2.5)  {\includegraphics[scale=0.8,valign=c,reflect]{Figures/Tubings/Functions/psi+-.pdf}};
\node at (2,-2.5)  {\includegraphics[scale=0.8,valign=c,reflect]{Figures/Tubings/Functions/psi++.pdf}};
\node at (0,-1.6)  {\includegraphics[scale=0.8,valign=c,reflect]{Figures/Tubings/Functions/psic+.pdf}};
\node at (3.4,0)  {\includegraphics[scale=0.8,valign=c,reflect]{Figures/Tubings/Functions/psi+c.pdf}};
\node at (-3.4,0)  {\includegraphics[scale=0.8,valign=c]{Figures/Tubings/Functions/psic+.pdf}};
\end{scope}
\end{tikzpicture}}
\label{equ:3site-geometry}
\eeq
The functions assigned to the nine vertices are those that sum up to the wavefunction. The functions on each edge connecting two vertices correspond to the source functions appearing in the differential equations of the functions assigned to these vertices. 
The tubings associated to these functions are obtained from the merger of adjacent tubes on each pair of attached vertex functions. The function associated to the face of the square is the source function appearing in the four differential equations for the functions assigned to its edges.

This geometric representation captures the fact that the differentials of functions involve the functions themselves, along with the functions associated to the attached objects of one lower co-dimension. It is straightforward to translate the equations for the three-site chain~\eqref{equ:psi3}--\eqref{equ:33} into this combinatoric language and see that they do indeed have the structure suggested by the geometry~\eqref{equ:3site-geometry}. 
For simplicity, we will not reproduce all these equations, but it is useful to note that the $A$-matrix takes the following schematic form
 \be
A =  \left[\includegraphics[valign=c,trim={0 5pt 0 5pt 0},scale=0.75]{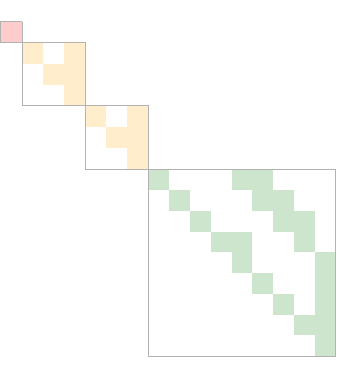}\right] ,
\label{equ:3site-Amatrix}
\ee
with the basis vector defined as 
\beq
\begin{aligned}
	\vec I = \ &( \red{ \psi_{\includegraphics[scale=0.6]{Figures/Tubings/Functions/psi00.pdf}}},\,\orange{ \psi_{\includegraphics[scale=0.6]{Figures/Tubings/Functions/psi-0.pdf}}},\,\orange{ \psi_{\includegraphics[scale=0.6]{Figures/Tubings/Functions/psi+0.pdf}}},\,\orange{ F_{\includegraphics[scale=0.6]{Figures/Tubings/Functions/psic0.pdf}}},\,\orange{ \psi_{\includegraphics[scale=0.6,reflect]{Figures/Tubings/Functions/psi-0.pdf}}},\,\orange{ \psi_{\includegraphics[scale=0.6,reflect]{Figures/Tubings/Functions/psi+0.pdf}}},\,\orange{ F_{\includegraphics[scale=0.6,reflect]{Figures/Tubings/Functions/psic0.pdf}}},\,\\ 
	&\ \green{\psi_{\includegraphics[scale=0.6,reflect]{Figures/Tubings/Functions/psi++.pdf}}},\,\green{\psi_{\includegraphics[scale=0.6]{Figures/Tubings/Functions/psi+-.pdf}}},\,\green{\psi_{\includegraphics[scale=0.6]{Figures/Tubings/Functions/psi++.pdf}}},\,\green{\psi_{\includegraphics[scale=0.6,reflect]{Figures/Tubings/Functions/psi-+.pdf}}},\\
	&\ \green{F_{\includegraphics[scale=0.6,valign=c,reflect]{Figures/Tubings/Functions/psi+c.pdf}}},\,\green{F_{\includegraphics[scale=0.6,valign=c,reflect]{Figures/Tubings/Functions/psic+.pdf}}},\,\green{F_{\includegraphics[scale=0.6,valign=c]{Figures/Tubings/Functions/psic+.pdf}}},\,\green{F_{\includegraphics[scale=0.6,valign=c]{Figures/Tubings/Functions/psi+c.pdf}}},\, \green{G_{\includegraphics[scale=0.6]{Figures/Tubings/Functions/psicc.pdf}}})\,.
\end{aligned}
\eeq
We clearly see that the matrix separates into one $1\times 1$ block, two $3\times 3$ blocks---with the same pattern of nonzero entries as in~\eqref{equ:2site-Amatrix}---and one $9\times 9$ block, which reflects the geometric structures in~\eqref{equ:3site-geometry}.

%\newpage
\item {\it Four-site graphs}\\[4pt]
We leave it as an exercise for the reader to show that the complete tubings of a four-site graph (chain or star) can be arranged into one point, three intervals, three squares and one cube. 
Shown below is the (partial) assignment of functions to the cube: 
\begin{align}
 \raisebox{-6.5em}{
	\begin{tikzpicture}[line width=2pt, scale=1.5]
	  \draw[color=Blue!50,dashed] (0.5,0.5) -- (0.5,2.5);
  \draw[color=Blue!50,dashed] (2.5,0.5) -- (0.5,0.5);
    \draw[color=Blue!50,dashed] (0,0) -- (0.5,0.5);
	\draw[fill=Blue,opacity=0.15] (0,0) rectangle (2,2);
		\draw[color=Blue] (0,0) rectangle (2,2);
    \draw[fill=Blue,opacity=0.15] (2,0) -- (2.5,0.5) -- (2.5,2.5) -- (2,2) --cycle;
     \draw[color=Blue] (2,0) -- (2.5,0.5) -- (2.5,2.5) -- (2,2) --cycle;
       \draw[fill=Blue,opacity=0.15] (0,2) -- (2,2) -- (2.5,2.5) -- (0.5,2.5) --cycle;
          \draw[color=Blue] (0,2) -- (2,2) -- (2.5,2.5) -- (0.5,2.5) --cycle;           
 \draw[fill, color=Blue] (0,2) circle (.65mm);	
 \draw[fill, color=Blue] (0,0) circle (.65mm);
 \draw[fill, color=Blue] (2,2) circle (.65mm);
 \draw[fill, color=Blue] (2,0) circle (.65mm);
 \draw[fill, color=Blue] (0+.5,2+.5) circle (.65mm);	
 \draw[fill, color=Blue] (0+.5,0+.5) circle (.65mm);
 \draw[fill, color=Blue] (2+.5,2+.5) circle (.65mm);
 \draw[fill, color=Blue] (2+.5,0+.5) circle (.65mm);
\node[inner sep=0pt] at (3.35,2.5)    {\includegraphics[scale=1]{Figures/Tubings/Functions/four/4topright}};
\node[inner sep=0pt] at (3.35,1.7)    {\includegraphics[scale=1]{Figures/Tubings/Functions/four/4rightedge}};
\node[inner sep=0pt] at (3.75-.2,1.3)    {\includegraphics[scale=1]{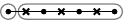}};
\node[inner sep=0pt] at (3.35,0.5)    {\includegraphics[scale=1]{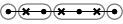}};
\node[inner sep=0pt] at (3.1,0.25)    {\includegraphics[scale=1]{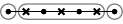}};
\node[inner sep=0pt] at (2.85,0)    {\includegraphics[scale=1]{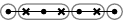}};
\node[inner sep=0pt] at (1.1,1.25)    {\includegraphics[scale=1]{Figures/Tubings/Functions/four/4vol}};
\node[inner sep=0pt] at (-.85,0)    {\includegraphics[scale=1]{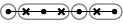}};
\node[inner sep=0pt] at (1,-0.2)    {\includegraphics[scale=1]{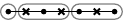}};
\node[inner sep=0pt] at (-.85,2)    {\includegraphics[scale=1]{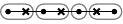}};
\node[inner sep=0pt] at (-.85,1)    {\includegraphics[scale=1]{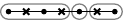}};
\node[inner sep=0pt] at (-.35,2.5)    {\includegraphics[scale=1]{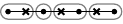}};
\node[inner sep=0pt] at (-0.65,2.25)    {\includegraphics[scale=1]{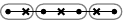}};
\node[inner sep=0pt] at (1.25,3.1)    {\includegraphics[scale=1]{Figures/Tubings/Functions/four/4topface}};
	  \draw[-stealth,line width =1pt] (2.8,1.3) -- (2.2,1.3);
      \draw[-stealth,line width =1pt] (1.25,2.95) -- (1.25,2.2);
	\end{tikzpicture}}
	\label{equ:cube}	
\end{align}
The vertex functions correspond to the fully-partitioned graph, i.e.~each individual tube only encircles a single vertex. The functions on the edges are obtained by the merger of adjacent tubes on each pair of attached vertex functions. Further mergers of tubes on the edge functions give the functions assigned to each face of the tube. Finally, the function assigned to the volume of the cube is the fully-encircled graph, which arises from the final mergers of tubes on the attached face functions.\footnote{Equivalently, we can view things in reverse, where the volume is associated to the tubing that fully encloses the graph. Moving to a facet then corresponds to pinching the tubing in one place, and further moving to edges and then vertices can be associated with further pinching of the tubes. In this way, the geometric shape captures the compatibilities between different ways of pinching tubings.}

The $A$-matrix for the four-site graph has the following pattern of nonzero entries
\beq
A =\left[\includegraphics[valign=c,scale=0.625]{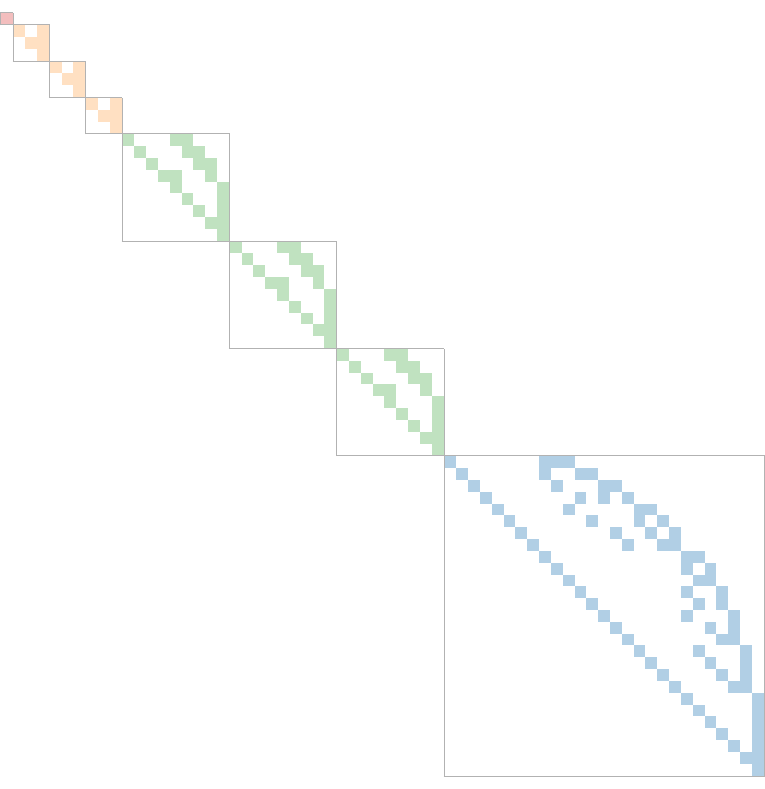}\right] .
\eeq
We clearly recognize the separation into independent blocks. The $3\times 3$ blocks are the same as for the two-site chain in~\eqref{equ:2site-Amatrix} and the $9\times 9$ blocks are the same as for the three-site chain in~\eqref{equ:3site-Amatrix}. The new $27 \times 27$ block reflects the cube in~\eqref{equ:cube}.
\end{itemize}
We now explain how to derive these differential equations directly from the geometry and its associated tubings.

%\newpage
\subsection{Kinematic Flow}
\label{ssec:flow}

The benefit of the graph tubing perspective that it allows for a simple algorithm to obtain the differential equations for arbitrary graphs (including loop integrands). We first describe this algorithm abstractly and then apply it to some illustrative examples.

\vskip 4pt
The starting point is the complete graph tubing associated to a ``parent function" of interest. Its differential follows from two extremely simple steps. 
\begin{itemize}
\item[1.] {\bf Activation}: The first step is the same as in~\cite{Arkani-Hamed:2023kig}. Each tube of the graph tubing gets ``activated" and becomes a letter in the differential equation. The coefficient of each letter is the function corresponding to the original graph times the sum of the twist parameters $\alpha_i$ associated to its enclosed vertices.

For example, consider the graph tubings $\includegraphics[scale=0.7]{Figures/Tubings/Functions/psi0.pdf}$ and $\includegraphics[scale=0.7]{Figures/Tubings/Functions/psic.pdf}$. According to the activation rule, the differentials of the associated functions are
\begin{align}
{\rm d}  \psi_{\includegraphics[scale=0.6]{Figures/Tubings/Functions/psi0.pdf}}   &\, = \, 
\Big(
\alpha_1   \, \includegraphics[scale=0.9,valign=c]{Figures/Tubings/two/Lap.pdf}   \, +\, \alpha_2     \includegraphics[scale=0.9,valign=c]{Figures/Tubings/two/Lbp.pdf}  \Big) \,  \psi_{\includegraphics[scale=0.6]{Figures/Tubings/Functions/psi0.pdf}}    \, , \label{equ:example1} \\ 
{\rm d} 
F_{\includegraphics[scale=0.6]{Figures/Tubings/Functions/psic.pdf}}  &\, =\, (\alpha_1 + \alpha_2)\, 
 \includegraphics[scale=0.9,valign=c]{Figures/Tubings/two/Lab.pdf}\, F_{\includegraphics[scale=0.6]{Figures/Tubings/Functions/psic.pdf}} \,. \label{equ:example2}
\end{align}
In~\eqref{equ:example1}, we have two letters arising from the two tubes of the tubing $\includegraphics[scale=0.7]{Figures/Tubings/Functions/psi0.pdf}$, while in~\eqref{equ:example2}, the coefficient is $\alpha_1+\alpha_2$ because two vertices are enclosed by the tube.

\item[2.] {\bf Merger}: If the graph tubing has tubes that are adjacent to each other, these tubes can merge to form a larger tube. 
The function associated to this new tubing will appear as a source function in the differential equation. This source function multiplies the {\it difference} of the two letters corresponding to the two tubes involved in the merger. The letter containing the cross on the merged edge appears with a plus sign.

This is best illustrated by examples.
Consider the graph tubing $\includegraphics[scale=0.7]{Figures/Tubings/Functions/psi-.pdf}$. 
The differential of the associated function is
\beq
{\rm d}\psi_{\includegraphics[scale=0.6]{Figures/Tubings/Functions/psi-.pdf}} \, = \, \Big(
\alpha_1   \, \includegraphics[scale=0.9,valign=c]{Figures/Tubings/two/Lam.pdf}   \, +\, \alpha_2     \includegraphics[scale=0.9,valign=c]{Figures/Tubings/two/Lbp.pdf}  \Big) \,  \psi_{\includegraphics[scale=0.6]{Figures/Tubings/Functions/psi-.pdf}} 
  \, +\, 
  \Big(
 \includegraphics[scale=0.9,valign=c]{Figures/Tubings/two/Lam.pdf}   \, -\,     \includegraphics[scale=0.9,valign=c]{Figures/Tubings/two/Lbp.pdf}  \Big)  \, F_{\includegraphics[scale=0.6]{Figures/Tubings/Functions/psic.pdf}} \ .
 \label{equ:example3}
 \eeq
The first term on the right-hand side comes from the activation rule, while the second term is the new source term arising from the merger.

In many cases there are multiple adjacent tubes. We simply add the effects of all possible pairwise mergers.
Consider, for example, the tubing $\includegraphics[scale=0.7,reflect]{Figures/Tubings/Functions/psi++.pdf}$.
The differential of the associated function is 
\beq
\begin{aligned}
{\rm d} \psi_{\includegraphics[scale=0.6,reflect]{Figures/Tubings/Functions/psi++.pdf}}  \ =\ &\Big( \alpha_1\,  
\includegraphics[scale=0.9,valign=c]{Figures/Tubings/three/LLam.pdf} + \alpha_2 \includegraphics[scale=0.9,valign=c]{Figures/Tubings/three/LLbpm.pdf} + \alpha_3 \includegraphics[scale=0.9,valign=c]{Figures/Tubings/three/LLcp.pdf} \Big)
\psi_{\includegraphics[scale=0.6,reflect]{Figures/Tubings/Functions/psi++.pdf}}          \\
&+\, \Big(\includegraphics[scale=0.9,valign=c]{Figures/Tubings/three/LLam.pdf}   \,-\,\includegraphics[scale=0.9,valign=c]{Figures/Tubings/three/LLbpm.pdf}  \Big)F_{\includegraphics[scale=0.6,reflect]{Figures/Tubings/Functions/psi+c.pdf}}  \\
&+\, \Big(\includegraphics[scale=0.9,valign=c]{Figures/Tubings/three/LLbpm.pdf}  \,-\,  \includegraphics[scale=0.9,valign=c]{Figures/Tubings/three/LLcp.pdf}   \Big)   F_{\includegraphics[scale=0.6,reflect]{Figures/Tubings/Functions/psic+.pdf}}\,.
\end{aligned} 
\label{equ:example4}
\eeq
The first line on the right-hand side comes from the activation rule, while the next two lines arise from the two mergers of adjacent tubes.
\end{itemize}
Applying these two simple steps will generate the differential equations for arbitrary tree graphs. It is worth noting that the procedure is dramatically simpler than the one in~\cite{Arkani-Hamed:2023kig}.

\subsection*{An Example}
The reader now has enough information to take any Feynman graph and use the rules above to derive the differential equations for its basis functions. 

\vskip 4pt
To illustrate the simplicity of the graphical rules, we present the example of the four-site chain. We have shown above that the basis functions (complete tubings) of this graph can be arranged into one point, three intervals, three squares and one cube. 
The (partial) assignment of functions to the cube was given in~\eqref{equ:cube}. For illustration purposes, let us pick one of the functions associated to a vertex of the cube, say $\includegraphics[scale=0.8,valign=c]{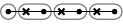}$. The differential of this function is
\beq
\begin{aligned}
{\rm d} \psi_{\includegraphics[scale=0.7,valign=c]{Figures/Tubings/Functions/four/4topright.pdf}} \ =\ & \Big(\alpha_1\,\includegraphics[scale=0.7,valign=c]{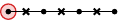}\,+\,\alpha_2\,\includegraphics[scale=0.7,valign=c]{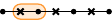}\,+\,\alpha_3\,\includegraphics[scale=0.7,valign=c]{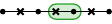}\,+\,\alpha_4\,\includegraphics[scale=0.7,valign=c]{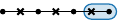}\Big)\psi_{\includegraphics[scale=0.7,valign=c]{Figures/Tubings/Functions/four/4topright.pdf}} \\
&+ \,\Big(\includegraphics[scale=0.7,valign=c]{Figures/Tubings/four/LLLap.pdf} \,-\,  \includegraphics[scale=0.7,valign=c]{Figures/Tubings/four/LLLbmp.pdf}\Big) F_{\includegraphics[scale=0.7,valign=c]{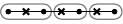}}\\
&+ \,\Big( \,\includegraphics[scale=0.7,valign=c]{Figures/Tubings/four/LLLbmp.pdf} \,-\,  \includegraphics[scale=0.7,valign=c]{Figures/Tubings/four/LLLcmp.pdf} \Big) F_{\includegraphics[scale=0.7,valign=c]{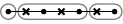}}\\
 &+ \,\Big(\, \includegraphics[scale=0.7,valign=c]{Figures/Tubings/four/LLLcmp.pdf} \,-\,  \includegraphics[scale=0.7,valign=c]{Figures/Tubings/four/LLLdm.pdf} \Big) F_{\includegraphics[scale=0.7,valign=c]{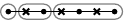}}\,.
\end{aligned}
\eeq
 We see that this equation has four letters, corresponding to the four tubes of the chosen parent function. These letters are created by the ``activation rule". Moreover, we see that the equation has three new source functions on the right-hand side corresponding to the three different ways of merging adjacent tubes on the parent function. Geometrically, these three functions correspond to the three edges attached to the chosen vertex on the cube. 
 The ``merger rule" explains how these source functions appear in the differential equation.
 
 \vskip 4pt 
 Next, we continue with one of the edge functions, say $ F_{\includegraphics[scale=0.7,valign=c]{Figures/Tubings/Functions/four/psic--.pdf}}$. 
 Its differential is 
 \beq
 \begin{aligned}
{\rm d} F_{\includegraphics[scale=0.7,valign=c]{Figures/Tubings/Functions/four/psic--.pdf}} \ = \ & \Big((\alpha_{1}+\alpha_2)\,\includegraphics[scale=0.7,valign=c]{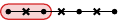} \,+\, \alpha_2\,\includegraphics[scale=0.7,valign=c]{Figures/Tubings/four/LLLcmp.pdf} \,+\, \alpha_3\,\includegraphics[scale=0.7,valign=c]{Figures/Tubings/four/LLLdm.pdf} \Big)F_{\includegraphics[scale=0.7,valign=c]{Figures/Tubings/Functions/four/psic--.pdf}}\\
 & + \, \Big(\includegraphics[scale=0.7,valign=c]{Figures/Tubings/four/LLLabp.pdf} \,-\,\includegraphics[scale=0.7,valign=c]{Figures/Tubings/four/LLLcmp.pdf}  \Big)G_{\includegraphics[scale=0.7,valign=c]{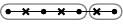}}\\
 & + \, \Big(\includegraphics[scale=0.7,valign=c]{Figures/Tubings/four/LLLcmp.pdf} \,-\,\includegraphics[scale=0.7,valign=c]{Figures/Tubings/four/LLLdm.pdf}  \Big)G_{\includegraphics[scale=0.7,valign=c]{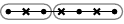}}\,,
 \end{aligned}
 \label{equ:4-edge}
 \eeq
 which has two source functions, corresponding to the two ways of merging adjacent tubes. Geometrically, these two functions correspond to the two faces that meet on the chosen edge. 
  
\vskip 4pt
Finally, we look at one of the functions associated to a face, $G_{\includegraphics[scale=0.7,valign=c]{Figures/Tubings/Functions/four/4topface.pdf}}$, 
whose differential is
 \beq
 \begin{aligned}
	{\rm d}G_{\includegraphics[scale=0.7,valign=c]{Figures/Tubings/Functions/four/4topface.pdf}} \ = \ \Big((\alpha_{1}+\alpha_2+\alpha_3) \, \includegraphics[scale=0.7,valign=c]{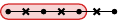} \,+\, \alpha_4\, \includegraphics[scale=0.7,valign=c]{Figures/Tubings/four/LLLdm.pdf}\Big)G_{\includegraphics[scale=0.7,valign=c]{Figures/Tubings/Functions/four/4topface.pdf}}&\\
	+\,\Big(\includegraphics[scale=0.7,valign=c]{Figures/Tubings/four/LLLabcp.pdf} \,-\,\includegraphics[scale=0.7,valign=c]{Figures/Tubings/four/LLLdm.pdf}\Big) H_{\includegraphics[scale=0.7,valign=c]{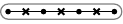}}&\,.
	 \end{aligned}
 \label{equ:4-face}
 \eeq
As expected, this involves a single source function coming from the merger of the two adjacent tubes on the parent function. Geometrically, this source function corresponds to the bulk volume of the cube attached to the chosen face. It is the common source function that appears in the differentials of all face functions.  Its differential is  
\beq
{\rm d} H_{\includegraphics[scale=0.7,valign=c]{Figures/Tubings/Functions/four/4vol.pdf}}  = (\alpha_1+\alpha_2+\alpha_3+\alpha_4) \, \includegraphics[scale=0.7,valign=c]{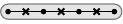}\,H_{\includegraphics[scale=0.7,valign=c]{Figures/Tubings/Functions/four/4vol.pdf}} \,.
\eeq
That is, it closes on itself as expected from the activation rule.

\vskip 4pt
One of the remarkable features of the differential is that it is  {\it local}, in the sense that it only depends on the merger of adjacent tubes. Therefore, nothing substantial changes when we consider the equations for arbitrarily complicated graphs. We simply continue merging adjacent tubes and write down the differentials using the activation and merger rules over and over. As a consequence, the differential is local in the geometry that organizes the functions.

\vskip 4pt
It is interesting to contrast the properties of the kinematic flow in this basis (which is natural from the time-integral perspective) with the basis considered in~\cite{Arkani-Hamed:2023kig}. A nice feature of that representation was that it began with the full wavefunction itself, and built the system of equations by introducing sources as needed until the system closed. While there is something canonical about this starting point, it obscures some of the underlying locality of the differential equations. In contrast, the compatibility relations between the various functions appearing in the system are most transparent in the time-integral basis considered here, because this is precisely the basis where locality in time is manifest.

\subsection*{Integrability} 

The kinematic flow rules only lead to consistent differential equations if they imply $\textrm{d}^2=0$. In~\cite{Arkani-Hamed:2023kig}, this was shown to be the case in a specific basis. Although this statement must be basis-independent, it is illuminating derive it in our new basis. Since the new kinematic flow rules are simpler, the derivation of $\ud^2 =0$ is also simpler.\footnote{Since the kinematic flow differential is ultimately a translation of the differential in the time-integral representation, it is guaranteed to be nilpotent. It is nevertheless illuminating to see directly in combinatorial language.}

\vskip 4pt
In the following, we present the action of $\ud^2$ on a generic function. By following the graphical rules iteratively, we conclude that the functions which appear in this equation will be those that can be produced by performing zero, one or two mergers. 
Each function is described by a list of its tubes, $\{T_i\}$. Associated to each tube is a letter $\ell_i$ and a twist parameter, $\alpha_i$. (If the tube encloses multiple vertices, the twist parameter is the sum of the twists for each enclosed vertex.) The merger of two tubes $T_i$ and $T_j$ produces a new function, with a new letter $\ell_{ij}=\ell_i+\ell_j$ and twist $\alpha_{ij} = \alpha_{i}+\alpha_j$. In~\cite{Arkani-Hamed:2023kig}, it was shown that  $\ud^2$ of any function whose differential involves such mergers produces terms proportional to
\beq
R_{ij} \equiv \ud \log \ell_i \wedge \ud \log \ell_j + \ud \log \ell_{ij} \wedge \ud \log\left(\frac{\ell_i}{\ell_j}\right) = 0\,.
\label{equ:3-letter}
\eeq
Except for the trivial case of no mergers, we find that
$\ud^2 =0$ because of the {\it three-letter relation} in~\eqref{equ:3-letter}.
\begin{itemize}
\item We first consider the case of zero mergers, which arises for tubings corresponding to functions that close on themselves, i.e.~fully-disconnected or fully-collapsed functions.  For example, this is the case for the functions  $\psi_{\includegraphics[scale=0.6]{Figures/Tubings/Functions/psi0.pdf}}$ and $F_{\includegraphics[scale=0.6]{Figures/Tubings/Functions/psic.pdf}}$ in~\eqref{equ:example1} and~\eqref{equ:example2}. The differential of such functions is
\beq
\ud F =  \bigg(\sum_{i} \alpha_i\, \ud \log \ell_i \bigg) F  \equiv \ud \Omega\, F\,,
\eeq
and hence $\ud^2 F = \ud \Omega \wedge \ud \Omega \, F = 0$.

\item Next, we look at the case of a single merger, with the simplest example being the function $\psi_{\includegraphics[scale=0.6]{Figures/Tubings/Functions/psi-.pdf}}$  in~\eqref{equ:example3}.
Let us generically write these functions as $F_{(i)(j)}$, where the subscripts denote the adjacent tubes $T_i$ and $T_j$. The merger of these tubes produces a new tube $T_{ij}$ and an associated new function $F_{(ij)}$. Using the flow rules, we then find that
\beq
\ud^2 F_{(i)(j)} = \pm \alpha_{ij} R_{ij} F_{(ij)} = 0 \,,
\eeq
where $R_{ij}$ was defined in~\eqref{equ:3-letter}. The plus or minus sign accounts for the fact that the cross on the line where the two tubes touch can be included in either $T_i$ of $T_j$ and the sign of the resulting term depends on this. However, since $R_{ij}=0$, the sign will be irrelevant and so, in the following, we will not keep track of this sign ambiguity.

\item Finally, we consider the case of two mergers. We first look at the example of three adjacent tubes, like for the function $\psi_{\includegraphics[scale=0.6,reflect]{Figures/Tubings/Functions/psi++.pdf}}$ in~\eqref{equ:example4}. We write generic functions of this type as $F_{(i)(j)(k)}$, where the adjacent tubes are $T_i$, $T_j$, $T_k$. Using the flow rules, we get
\beq\label{equ:d2Fijk}
\ud^2 F_{(i)(j)(k)} = \alpha_{ij} R_{ij} F_{(ij)(k)} +  \alpha_{jk} R_{jk} F_{(i)(jk)}  + \alpha_{ijk} (R_{ij}+R_{jk}) F_{(ijk)} = 0\,,
\eeq
which again vanishes because of the vanishing three-letter relations.

A similar analysis for a function with four adjacent tubes leads to 
\beq
\begin{aligned}
\ud^2 F_{(i)(j)(k)(l)} &= \alpha_{ij} R_{ij} F_{(ij)(k)(l)} +  \alpha_{jk} R_{jk} F_{(i)(jk)(l)} + \alpha_{kl} R_{kl} F_{(i)(j)(kl)}  \\
&\quad + \alpha_{ijk} (R_{ij}+R_{jk}) F_{(ijk)(l)} + \alpha_{jkl}(R_{jk}+R_{kl}) F_{(i)(jkl)} \\
&\quad + \alpha_{ij} \alpha_{kl} S_{ij,kl} F_{(ij)(kl)} = 0\,,
\end{aligned}
\eeq
where 
\begin{align}\label{equ:d2Fijkl}
S_{ij,kl} \equiv \textrm{d}\log\left(\frac{\ell_i}{\ell_j}\right)\wedge \textrm{d}\log\left(\frac{\ell_k}{\ell_l}\right)+\textrm{d}\log\left(\frac{\ell_k}{\ell_l}\right)\wedge \textrm{d}\log\left(\frac{\ell_i}{\ell_j}\right) =0\,.
\end{align}
The term proportional to $S_{ij,kl}$ arises from two pairs of tubes merging, which can happen in two ways and therefore vanishes because of the anti-symmetry of the wedge product.
\end{itemize}
Since the kinematic flow rules act locally, the above observations are sufficient to prove the vanishing of ${\rm d}^2$ on any function.

\subsection{Boundary Conditions}
\label{ssec:BC}

A practical benefit of the new basis used here is that it automatically puts the differential equations into a form that is easy to solve. In this section, we will explain this feature and describe the boundary conditions  that we impose to fix the solutions.
An explicit example is given in Appendix~\ref{app:solving}.

\vskip 4pt
One simplification comes from the fact that the differential equations separate into independent sectors graded by the number~$r$ of ``disconnected edges" (i.e.~tubings with  un-circled crosses, or graphs with non-time-ordered propagators). These sectors can be solved independently and the full solution for an $n$-site graph can be written as 
\be
	\psi_n =\sum_{r=0}^e \sum_{m=1}^{{}^e C_r} 
	c_{r,m} \psi_n^{(r,m)}\,,
\label{equ:sum}
\ee
where $c_{r,m}$ are constant coefficients that will be fixed by the boundary conditions.
 Moreover, the solution for the sectors with $r >0$ can obtained iteratively by recycling the solutions of smaller graphs (i.e.~graphs with fewer edges).  From the bulk perspective, this is simply leveraging the fact that the disconnected internal edges separate the solution into a product of lower-point solutions. The new information is therefore contained in the fully-connected graphs with $r=0$.

\vskip 4pt
Within each sector, the differential equations are most easily solved ``inside out". First, we solve for the function associated to the bulk of the zonotope of interest. Since only this function itself appears in its differential, we can easily integrate the corresponding differential equation to get a power-law solution (with an unfixed integration constant).  
Using this solution as a source, we next integrate the differential equations for the functions associated to the facets of the zonotope, whose differential involves themselves and the function associated to the bulk volume.
Integrating the power-law solution generally leads to hypergeometric functions. This procedure is continued until one obtains the solutions for the vertex functions.\footnote{It is worth noting that this organization also makes clear that the completely time-ordered contributions will be the most complicated functions, since they will correspond to the highest-dimension geometry, and hence the most nested integrations. In the de Sitter limit, this can be made precise, where the fully time-ordered contribution will lead to the highest-weight polylogarithmic function appearing in the wavefunction, while the disconnected contributions give products of functions with lower transcendental weight.} These functions are then added together to obtain the solution in~\eqref{equ:sum}.

\vskip 4pt 
To select the solution of interest, we must impose suitable {\it boundary conditions}. 
As explained in~\cite{Arkani-Hamed:2018kmz,Arkani-Hamed:2023kig}, one condition is the absence of any folded singularities (assuming the Bunch--Davies vacuum).\footnote{Folded singularities occur in a configuration where the sum of some external energies adds up to an internal energy. The limit corresponds to ``folding" the kinematic polygon to make external edges lie along an internal line.}
In fact, it is easy to see from the time-integral representation that these folded singularities cannot appear in any of the basis functions.
Solving the system from the inside out, each function satisfies a first-order equation, and so has a single free integration constant associated to it. Forbidding folded singularities fixes all of these coefficients except that of the lowest-level source function associated to the geometries' volume (whose equations don't contain any folded letters). Thus, we are left with one free coefficient for each zonotope. 

\vskip 4pt
A Feynman graph with $e$ edges will have $2^e$ free coefficients after imposing the absence of all folded singularities, one for each of the disconnected geometries. To fix these coefficients, we impose the correct factorization of the wavefunction on all partial energy singularities~\cite{Arkani-Hamed:2017fdk,Baumann:2020dch,Goodhew:2021oqg}. These boundary conditions will mix the different sectors in the differential equation together, since they fixed the relative normalizations $c_{r,m}$ of the solutions in (\ref{equ:sum}). Although there can be fewer than $2^e$ partial energy singularities, the functional form of the residue always imposes enough constraints to uniquely fix the solution.

\vskip 4pt
To make this discussion less abstract, consider an $(n+m)$-point tree graph.\footnote{The same arguments can be applied at loop level, the only complication being that the form of the coefficient of the singularity will be slightly more complicated (but still related to lower-point information)~\cite{Arkani-Hamed:2017fdk,Melville:2021lst,Baumann:2021fxj}.} For concreteness, let us focus on the factorization singularity where $n$ of the external energies add up to minus the internal energy $Y$ in such a way that splits the graph into two pieces, one with $n$ vertices, and one with~$m$. In this limit, $E_n \equiv X_1+\cdots +X_n+Y\to 0$, the wavefunction coefficient factorizes as
\be
\lim_{E_n \to 0}\psi_{n+m}
\propto \frac{1}{E_n^\gamma}A_{n} \times \Big[ \psi_m(X,-Y) - \psi_m(X, Y)  \Big]\, ,
\label{equ:factorization}
\ee
where $\gamma$ is the order of the singularity ($\gamma \to 0$ should be interpreted as a branch point) and $A_n$ is the scattering amplitude associated to the subgraph whose energy is conserved. The second factor is a ``shifted wavefunction coefficient"~\cite{Arkani-Hamed:2017fdk,Baumann:2020dch,Goodhew:2021oqg}, which depends on the energies associated to the graph with $m$ vertices, collectively denoted as $X$, along with $Y$. Each term in this shifted wavefunction coefficient can be understood as the sum over $2^{m-1}$ zonotopes. This limit therefore fixes $2\times2^{m-1}$ of the remaining free coefficients. As a practical matter, it is simplest to set $n=1$, so that we can fix all coefficients by imposing a single factorization limit.\footnote{Note that we have described the boundary conditions for the wavefunction coefficients. Since correlation functions satisfy the same differential equations, the difference between the two objects only lies in boundary conditions. To extract correlators as opposed to wavefunction coefficients, one only needs to enforce the factorization appropriate for a correlator.}

\subsection{Emergent Time}

Starting from the representation of the cosmological wavefunction as an integral over time, we saw that the differential equations which describe its kinematic dependence can be given a combinatorial and geometric interpretation.
Since our aim in cosmology is to infer the properties of time evolution from static observations at the end of inflation, it is very tempting to consider telling the story backwards, to see how time evolution emerges from the flow in kinematic space~\cite{Arkani-Hamed:2023bsv}. 
This more radical viewpoint is most natural in the basis of functions described in this paper because the objects are arranged into building blocks that naturally arise from the time-integral representation. Here, we provide a sketch for such a narrative.

\vskip 4pt
We begin by considering the kinematics of static correlation functions defined on the spatial boundary of a spacetime. In Fourier space, an $n$-point function depends on $n$ momentum vectors, which form a closed $n$-gon as a consequence of translation invariance.
The invariant kinematic data can be thought of as the shapes of the various subpolygons of the $n$-gon. 
In this space of kinematics, there are two special kinematic configurations. First, we can arrange the kinematics so that a subpolygon has zero area, corresponding  to the limit where some of the external momenta are collinear with an internal momentum (a so-called ``folded" limit). 
Second, if we complexify momenta, we can reach a configuration where the perimeter of a (sub)polygon vanishes. In principle, the correlation function (or wavefunction coefficient) could be singular in either of these limits. For a fixed triangulation of the kinematic $n$-gon, these kinematic configurations can also be represented by tubings of a marked graph.

\vskip 4pt
A natural question to ask is how the correlations change as we vary the kinematic data. The answer to this question is a differential equation. Hence, the goal is to find a structure in the space of kinematics that produces a differential equation. We expect that the special kinematic limits will appear as letters in this differential equation. Since these letters can be associated to tubings of marked graphs, it is natural to search for a structure involving graph tubings.  Interestingly, the full wavefunction has such a combinatorial construction as an integral over a rational function associated to graph tubings~\cite{Arkani-Hamed:2017fdk,Baumann:2024mvm}. In~\cite{Arkani-Hamed:2023bsv}, the resulting structure was related to time evolution, with the kinematic flow capturing some aspects of bulk physics. However, it remained a challenge to fully ``integrate in" time. The new perspective explored here provides a plausible avenue to make such an explicit connection.

\vskip 4pt
Since the various kinematic limits are associated to tubings of a marked graph, it is natural to search for an organizing principle that captures the relations between these tubings.
For a tree-level graph with $e$ edges, we construct $4^{e}$ ``complete tubings" of the marked graph and $2e^2+2e + 1$ tubes.  These become respectively the functions and letters in a set of coupled differential equations.  The complete tubings (basis functions) can be graded by the number of exposed crosses, $r=0,\cdots, e$. Only functions with the same grading mix in the differential equations. Graph tubings with the same grade can be arranged geometrically and assigned to the vertices, edges, faces, etc., of a polytope. Functions with the largest number of tubes  (those encircling only single vertices) are in one-to-one correspondence with the vertices of this geometry. The edges that connect these vertices correspond to the tubings that arise when we merge together various tubes in pairs, and reflect the compatibilities of this operation. In the same way, faces that meet on edges correspond to further mergers, and so on. 
The ``edge functions" become sources in the differential equations for the connected ``vertex functions". In the differential for the edge functions, the functions associated to facets appear, and the pattern continues to lower codimension objects.

\vskip 4pt
The arrangement of graded complete tubings into points, intervals, squares and (hyper)cubes geometrizes the flow in kinematic space. Taking derivatives with respect to the external energies creates differential equations with a pattern of letters and coefficient functions that precisely matches the way the tubings are connected into these geometrical shapes.
This requires a precise rule for the differential, but this is sufficiently constrained by integrability 
that there are a limited number of possibilities (it would be extremely interesting to prove that the merger rule described here is unique).
A beautiful feature of this perspective is that
tubes can only ``grow" by the merger of adjacent tubes until the entire graph is encircled by a single tube. The system starts with ``small singularities" that monotonically become larger when we take derivatives in kinematic space. This is how bulk time evolution is encoded in the boundary correlators. Indeed, there is now a clear correspondence between these growing tubes and the collapse of time-ordered propagators used in the bulk perturbation theory. In a sense, the monotonic evolution forward in time in the bulk has been translated into a monotonic growth towards larger tubings in the boundary kinematics.
Perhaps the most striking feature of this construction is that the boundary combinatoric description is {\it local} in the auxiliary geometric space that organizes the graph tubings. 

\vskip 4pt
Of course, this telling of the story is somewhat revisionist. We began with the bulk time perspective, and so it is somewhat of a cheat to obfuscate it and then imagine re-introducing time.   
Moreover, although it is fairly natural, we do not have an orthogonal justification for {\it why} a boundary organism should be interested in the geometry of all possible graph tubings, since this is an auxiliary structure beyond pure kinematics. 
Perhaps more importantly, we also do not have a first-principles explanation why the object of interest (the wavefunction) should correspond to summing together the functions associated to all the vertices of the geometry. Such a justification is important, not least because it is this fact that ultimately links together the disparate geometric pieces.
Relatedly, it would be desirable to have some other principle that defines/determines the differential of the functions, perhaps related to the geometry itself more directly.
In a sense, the deepest fact is the nontrivial compatibility between the wavefunction-centric construction of~\cite{Arkani-Hamed:2023kig} and the bulk time-integral perspective that was our starting point here, through a change of basis. A true emergence of time would therefore  explain this change of basis along with the other structures that we have observed.

%%%%%%%%%%%%%%%%%%%%%%%%%%%%%%%%%%%%%%%%%%%%%%%%%%%%%%
\section{Conclusions and Outlook}
\label{sec:conclusions}

A key feature of cosmological correlators is that their spatial dependence encodes dynamical properties of the bulk time evolution. In particular, consistent evolution is reflected in special properties of the differential equations obeyed by the boundary correlators.  Remarkably, in the case of a conformally coupled scalar field in power-law FRW spacetimes, these differential equations describe a flow in kinematic space~\cite{Arkani-Hamed:2023kig}.  In this paper, we have described a geometric origin for this flow which elucidates many of its properties.

\vskip4pt
It is interesting that the functions which arise in cosmology have a dynamical interpretation both in spacetime and on the boundary. 
The nontrivial compatibility of these two descriptions imbues cosmological observables with a large amount of structure which makes them special mathematical objects. We have seen one important facet of this compatibility: The locality and causality of bulk time evolution---as expressed by the collapsing of time-ordered propagators when taking derivatives---translates into locality of the kinematic flow in the abstract  space of graph tubings. Insights also flow in the other direction. By phrasing the problem combinatorially, we have been able to see structures that are obfuscated in the time-integral representation. 

\vskip 4pt
Looking forward, more fully understanding these geometric structures will surely lead to new insights. We end by describing some natural directions for further investigation:
\begin{itemize}
\item So far, the pattern of the kinematic flow has only been identified for conformally coupled scalars. It is natural to wonder whether it is special to this setup.
Ultimately, the origin of kinematic flow is tied to the collapsing of time-ordered propagators. This causal structure persists for massive fields, as well as for ``unparticles" (bulk CFT primaries; see e.g.~\cite{Pimentel:2025rds} in the context of cosmology). We will discuss the resulting pattern in future work.

\item Computing loop momentum integrals for cosmological correlators remains an important challenge, since the measure of the momentum integration changes the analytic structure of the integrand. The strategy outlined in this paper (as well as techniques based on hyperplane arrangements~\cite{Arkani-Hamed:2023kig}), does not immediately apply in those cases. Nevertheless, we have found a simple geometric structure underlying the differential equations for the loop integrands, and it would be interesting if this could be leveraged for the evaluation of loop integrals. In particular, the separation into independent sectors may suggest new strategies for renormalizing these loops.

\item From the mathematical side, many of the structures uncovered are deserving of further study and deeper understanding. In particular, the exterior derivative operation defines a notion of cohomology on directed acyclic graphs, and it would be interesting to explore whether there is any utility of this construction. More modestly, it would be nice to fully characterize the zonotope geometry for arbitrary loop integrands, and to understand the combinatorics of tubings of marked graphs in a more general setting.

\item We have only presented results for individual Feynman graphs. However, there is almost surely more structure in the sum over graphs in colored/flavored theories. The
simplest theory where individual graphs combine together irreducibly is the ${\rm tr}\hs \phi^3$ theory of colored scalars. Indeed, there is an interesting pattern of ``shared functions"~\cite{Arkani-Hamed:2023kig}. In Appendix~\ref{app:trp3}, we briefly give a more geometric meaning to this pattern. In particular, we show that ``nested" associahedra naturally describe the connections between basis functions for all exchange channels. In the future, we hope to further understand this geometry and to study how it can help solve the differential equations for the sum over graphs.

\item In~\cite{Arkani-Hamed:2024jbp}, a geometry---the ``cosmohedron"---was discovered, which describes the wavefunction of ${\rm tr}\hs \phi^3$ theory in flat space. It would be amazing if such a geometry also exists for the wavefunction in a cosmological background. This would be a generalization of the cosmohedron {\it after} performing the time integrals for each bulk vertex. Since these time integrals lead to a more complicated analytic structure of the wavefunction (not just poles like in flat space), we don't expect this object to be a polytope. A more modest first goal would be to understand the interplay between the geometric structures described here and the cosmohedron, viewed as an integrand for the FRW wavefunction. A further ambition is to combine these insights  with those of~\cite{Arkani-Hamed:2019mrd} to define stringy correlators in cosmology.
\end{itemize}

\noindent
In some sense, the most remarkable feature of the geometric picture we have described in this paper is that there is a geometry at all. A random collection of master integrals has no reason to have an ordering suitable to be realized geometrically. Of course, the special feature in this case is that the functions themselves are intrinsically tied to {\it time}. The combinatorics on the boundary is inherited from the bulk dynamics, but 
it can also be given an autonomous definition in terms of graph tubings.
Since we are studying a ``dual"  geometry of cosmological time integration, 
it is natural to suspect that more fully understanding this geometry will allow us to more directly lift insights from flat space into the cosmological setting. The most fantastical hope would be that there are some local microphysical laws that can be defined in this geometric space from which all of the features we have seen (including the bulk spacetime) derive.

 \vspace{0.2cm}
 \paragraph{Acknowledgments} DB, AJ, HL and GLP thank Nima Arkani-Hamed and Aaron Hillman for the collaboration that led to the discovery of the kinematic flow. 
 We are grateful to Nima Arkani-Hamed, Shounak De, Livia Ferro, Callum Jones, Gordon Lee, Tomasz Lukowski, Jiajie Mei, Ian Moult, Enrico Pajer, Shruti Paranjape, Facundo Rost, Andrzej Pokraka, Kamran Salehi Vaziri, Bernd Sturmfels and Jaroslav Trnka for helpful discussions.
 DB and HG wish to thank Claudia Fevola and Martina Juhnke-Kubitzke for many insightful discussions on the combinatorics of cosmological integrals.
 
 \vskip 4pt
The research of DB is funded by the European Union (ERC,  \raisebox{-2pt}{\includegraphics[height=0.9\baselineskip]{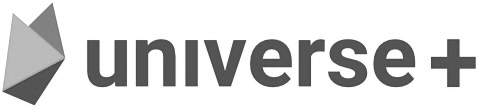}}, 101118787). Views and opinions expressed are, however, those of the author(s) only and do not necessarily reflect those of the European Union or the European Research Council Executive Agency. Neither the European Union nor the granting authority can be held responsible for them. DB is further supported by a Yushan Professorship at National Taiwan University funded by the Ministry of Education (MOE) NTU-112V2004-1. DB thanks the Max-Planck-Institute for Physics (MPP) in Garching for its hospitality while some of this work was being performed. He is grateful to the Alexander von Humboldt Stiftung and the Carl Friedrich von Siemens Stiftung for supporting his visits to the~MPP.
\vskip 4pt
HG is supported by a Postdoctoral Fellowship at National Taiwan University funded by the National Science and Technology Council (NSTC) 113-2811-M-002-073.
\vskip 4pt
AJ is supported in part by DOE (HEP) Award DE-SC0025323 and by the Kavli Institute for Cosmological Physics at the University of Chicago. 

\vskip 4pt
GLP presented some of these results at the workshop ``What is Particle Theory?" at KITP, Santa Barbara, as well as in seminars at UCLA and IAS (Princeton), and he is grateful for many interesting discussions and feedback from the participants. 
GLP thanks the KITP, as well as IAS Princeton for their hospitality and support.
GLP and TW are supported by Scuola Normale, by INFN (IS GSS-Pi), and by the ERC (NOTIMEFORCOSMO, 101126304). 
Views and opinions expressed are, however, those of the author(s) only and do not necessarily reflect those of the European Union or the European Research Council Executive Agency. 
Neither the European Union nor the granting authority can be held responsible for them. GLP is further supported by a Rita-Levi Montalcini Fellowship from the Italian Ministry of Universities and Research (MUR), as well as under contract 20223ANFHR (PRIN2022). This research was supported in part by grant NSF PHY-2309135 to the Kavli Institute for Theoretical Physics (KITP). 

%%%%%%%%%%%%%%%%%%%%%%%%%%%%%%%%%%%%%
\newpage
\appendix
%%%%%%%%%%%%%%%%%%%%%%%%%%%%%%%%%%%%%
\section{Solving the Differential Equations}
\label{app:solving}

In this appendix, we describe the process of solving the differential equations for the two-site chain. Recall that the relevant differential equations  are
\begin{align}
{\rm d}  \psi_{\includegraphics[scale=0.6]{Figures/Tubings/Functions/psi0.pdf}}   &\, = \, \Big(
\alpha_1   \, \includegraphics[scale=0.9,valign=c]{Figures/Tubings/two/Lap.pdf}   \, +\, \alpha_2     \includegraphics[scale=0.9,valign=c]{Figures/Tubings/two/Lbp.pdf}  \Big) \,  \psi_{\includegraphics[scale=0.6]{Figures/Tubings/Functions/psi0.pdf}} \ ,  \label{Aequ:4-3}\\
{\rm d}\psi_{\includegraphics[scale=0.6]{Figures/Tubings/Functions/psi+.pdf}}  &\, = \, \Big(
\alpha_1   \, \includegraphics[scale=0.9,valign=c]{Figures/Tubings/two/Lap.pdf}   \, +\, \alpha_2     \includegraphics[scale=0.9,valign=c]{Figures/Tubings/two/Lbm.pdf}  \Big) \,  \psi_{\includegraphics[scale=0.6]{Figures/Tubings/Functions/psi+.pdf}} 
  \, +\, 
  \Big(\includegraphics[scale=0.9,valign=c]{Figures/Tubings/two/Lbm.pdf}   \, -\,     \includegraphics[scale=0.9,valign=c]{Figures/Tubings/two/Lap.pdf}  \Big)  \, F_{\includegraphics[scale=0.6]{Figures/Tubings/Functions/psic.pdf}}   \,, \label{Aequ:4-4}\\
{\rm d}\psi_{\includegraphics[scale=0.6]{Figures/Tubings/Functions/psi-.pdf}} &\, = \, \Big(
\alpha_1   \, \includegraphics[scale=0.9,valign=c]{Figures/Tubings/two/Lam.pdf}   \, +\, \alpha_2     \includegraphics[scale=0.9,valign=c]{Figures/Tubings/two/Lbp.pdf}  \Big) \,  \psi_{\includegraphics[scale=0.6]{Figures/Tubings/Functions/psi+.pdf}} 
  \, +\, 
  \Big(
 \includegraphics[scale=0.9,valign=c]{Figures/Tubings/two/Lam.pdf}   \, -\,     \includegraphics[scale=0.9,valign=c]{Figures/Tubings/two/Lbp.pdf}  \Big)  \, F_{\includegraphics[scale=0.6]{Figures/Tubings/Functions/psic.pdf}}  \,, \label{Aequ:4-5}\\[2pt]
{\rm d} 
F_{\includegraphics[scale=0.6]{Figures/Tubings/Functions/psic.pdf}}  &\, = \, (\alpha_{1}+\alpha_2)\, 
 \includegraphics[scale=0.9,valign=c]{Figures/Tubings/two/Lab.pdf}\ F_{\includegraphics[scale=0.6]{Figures/Tubings/Functions/psic.pdf}} \,.
\label{Aequ:4-6}
\end{align}
In the following, we will solve this system sector by sector, and then combine all the vertex functions $\psi_i$, with the correct relative normalizations, to obtain the full solution.

\subsection{Power-Law Solutions}
The  solutions of (\ref{Aequ:4-3}) and (\ref{Aequ:4-6})
are simple power laws 
\begin{align}
	\psi_{\includegraphics[scale=0.6]{Figures/Tubings/Functions/psi0.pdf}} &\,=\, c_{\includegraphics[scale=0.6]{Figures/Tubings/Functions/psi0.pdf}}\,  (X_1+Y)^{\alpha_1}(X_2+Y)^{\alpha_2}\,,\label{vertexfunc}  \\
	F_{\includegraphics[scale=0.6]{Figures/Tubings/Functions/psic.pdf}}  &\,=\, c_{\includegraphics[scale=0.6]{Figures/Tubings/Functions/psic.pdf}} \,  (X_1+X_2)^{\alpha_{1}+\alpha_2}\,, \label{equ:F-sol}
\end{align}
where $c_{\includegraphics[scale=0.6]{Figures/Tubings/Functions/psi0.pdf}}$ and  $c_{\includegraphics[scale=0.6]{Figures/Tubings/Functions/psic.pdf}}$ are unfixed integration constants.   The relative size of the two constants will later be fixed by imposing the correct factorization limit.

\subsection{Inhomogeneous Solution}
The solution (\ref{equ:F-sol}) then enters as a source in (\ref{Aequ:4-4}) and (\ref{Aequ:4-5}).
For example, equation \eqref{Aequ:4-4} can be expressed as two separate ODEs as
\begin{align}
	\bigg(\partial_{X_1} -\frac{\alpha_1}{X_1+Y}\bigg)\psi_{\includegraphics[scale=0.6]{Figures/Tubings/Functions/psi+.pdf}} &=  -\, \frac{c_{\includegraphics[scale=0.6]{Figures/Tubings/Functions/psic.pdf}}}{X_1+Y}\,(X_1+X_2)^{(\alpha_1+\alpha_2)}\,,\\
	\bigg(\partial_{X_2} -\frac{\alpha_2}{X_2-Y}\bigg)\psi_{\includegraphics[scale=0.6]{Figures/Tubings/Functions/psi+.pdf}} &=  +\, \frac{c_{\includegraphics[scale=0.6]{Figures/Tubings/Functions/psic.pdf}}}{X_2-Y}\, (X_1+X_2)^{(\alpha_1+\alpha_2)}\,,
\end{align}
where we have substituted the source term for $F_{\includegraphics[scale=0.6]{Figures/Tubings/Functions/psic.pdf}}$. 
To solve these equations, it is useful to write
\begin{align}
	\psi_{\includegraphics[scale=0.6]{Figures/Tubings/Functions/psi+.pdf}} \equiv (X_1+X_2)^{\alpha_{1}+\alpha_2}\,\big(f(\xi) +  h(\xi) \big)\,,\label{psi+fh}
\end{align}
where $\xi \equiv (X_2-Y)/(X_1+Y)$ and $h(\xi)$ is a solution to the homogeneous equation.
The equations for the inhomogeneous solution become
\begin{align}
	\left(\xi\partial_\xi +\alpha_1 - \frac{\alpha_{1}+\alpha_2}{1+\xi}\right)f&= \,c_{\includegraphics[scale=0.6]{Figures/Tubings/Functions/psic.pdf}}\,, \label{equ:ODE1}\\
	\left(\xi\partial_\xi-\alpha_2+ \frac{\alpha_{1}+\alpha_2}{1+\xi^{-1}}\right) f&= \,c_{\includegraphics[scale=0.6]{Figures/Tubings/Functions/psic.pdf}}\,.\label{equ:ODE2}
\end{align}
We solve this using a series expansion, in two regimes separately: $|\xi|\le 1$ and $|\xi|\ge 1$.

\begin{itemize}
\item For $|\xi| \le 1$, we start with the ansatz
\begin{align}
	f_{<}(\xi) = \sum_{m= 0}^\infty a_m \xi^m\,,
\end{align}
which is regular at $\xi = 0$.
This is a solution of (\ref{equ:ODE1}) and (\ref{equ:ODE2}) if
\beq
\begin{aligned}
 a_0 &= -\frac{c_{\includegraphics[scale=0.6]{Figures/Tubings/Functions/psic.pdf}}}{\alpha_2}\,,\\
 a_m &= -\,c_{\includegraphics[scale=0.6]{Figures/Tubings/Functions/psic.pdf}}\frac{\alpha_{1}+\alpha_2}{\alpha_1\alpha_2} \frac{(\alpha_1)_m}{(1-\alpha_2)_m}(-1)^m \quad (m\ge 1)\,.
 \end{aligned}
 \eeq
The solution can be expressed as
\begin{align}
	f_{<}(\xi) = c_{\includegraphics[scale=0.6]{Figures/Tubings/Functions/psic.pdf}}\left[\frac{1}{\alpha_1}-\frac{\alpha_{1}+\alpha_2}{\alpha_1\alpha_2}\sum_{m= 0}^\infty \frac{(\alpha_1)_m}{(1-\alpha_2)_m}(-\xi)^m\right] \quad (|\xi|\le 1)\,.\label{fless}
\end{align}
This series converges absolutely at $|\xi|=1$ (or $X_1+X_2=0$) for $\alpha_1+\alpha_2<0$. For $\alpha_1+\alpha_2\ge 0$, the branch cut at $|\xi|=1$ cancels with the prefactor $(X_1+X_2)^{\alpha_1+\alpha_2}$ in \eqref{psi+fh}, so that the limit remains well defined. In this specific example, the series solution in \eqref{fless} could also be written in terms of the Gauss hypergeometric function ${}_2 F_1$, which could then be analytically continued to the regime $|\xi| > 1$. In other examples, however, we won't have this luxury, so we will instead continue working with the explicit series solution.

\item For $|\xi| \ge 1$,  we use as the ansatz a series expansion around $\xi^{-1} = 0$:
\begin{align}
	f_{>}(\xi) = \sum_{m= 0}^\infty \tilde a_m \frac{1}{\xi^m}\,.
\end{align}
The solution can be found in a similar manner and is given by
\begin{align}
	f_{<}(\xi) = -c_{\includegraphics[scale=0.6]{Figures/Tubings/Functions/psic.pdf}}\left[\frac{1}{\alpha_2}-\frac{\alpha_{1}+\alpha_2}{\alpha_1\alpha_2}\sum_{m= 0}^\infty \frac{(\alpha_2)_m}{(1-\alpha_1)_m}\frac{1}{(-\xi)^m}\right] \quad (|\xi|\ge 1)\,,\label{fgtr}
\end{align}
which differs from \eqref{fless} by the exchange $\alpha_1\leftrightarrow \alpha_2$, $\xi\leftrightarrow 1/\xi$, and an overall minus sign.
\end{itemize}

\subsection{Homogeneous Solution}

The homogeneous solution is
\begin{align}
	h(\xi) = c_{\includegraphics[scale=0.6]{Figures/Tubings/Functions/psi+.pdf}} \, \frac{\xi^{\alpha_2}}{(1+\xi)^{\alpha_1+\alpha_2}} \,.
\label{sol:homo}
\end{align}
We will fix the homogeneous solution by imposing the absence of a folded singularity at $\xi=0$ (or $X_2-Y=0$) and continuity at $|\xi|=1$.
\begin{itemize}
\item For $|\xi|\le 1$, the solution (\ref{sol:homo}) has a branch point at $\xi=0$. 
Since the inhomogeneous solution $f_<(\xi)$ is regular at $\xi=0$, we must set $c_{\includegraphics[scale=0.6]{Figures/Tubings/Functions/psi+.pdf}} = 0$ to forbid the appearance of a folded singularity. The final solution, for $|\xi|\le 1$, then is 
\begin{align}
	\psi_{\includegraphics[scale=0.6]{Figures/Tubings/Functions/psi+.pdf}} =  c_{\includegraphics[scale=0.6]{Figures/Tubings/Functions/psic.pdf}}\,(X_1+X_2)^{\alpha_{1}+\alpha_2}\left[\frac{1}{\alpha_1}-\frac{\alpha_{1}+\alpha_2}{\alpha_1\alpha_2}\sum_{m= 0}^\infty \frac{(\alpha_1)_m}{(1-\alpha_2)_m}(-\xi)^m\right] \ \  (|\xi|\le 1)\,,\label{Fsol1}
\end{align}
which is the same as $(X_1+X_2)^{\alpha_1+\alpha_2}f_<(\xi)$.

\item For $|\xi|\ge 1$, we avoid the branch point at $\xi=0$, so we can, in principle, add a nonzero homogeneous piece,  $f_>(\xi) + h(\xi)$. To fix the homogeneous solution, we impose continuity at $\xi = -1$. This gives
\begin{align}
	h(\xi) = c_{\includegraphics[scale=0.6]{Figures/Tubings/Functions/psic.pdf}}\,\frac{\Gamma(\alpha_1)\Gamma(\alpha_2)}{\Gamma(-\alpha_1-\alpha_2)}\, \frac{\xi^{\alpha_2}}{(1+\xi)^{\alpha_1+\alpha_2}} \ \ (|\xi|\ge 1)\,.
\end{align}
We see that the size of the homogeneous solution,  for $|\xi|\ge 1$, is determined by the integration constant $c_{\includegraphics[scale=0.6]{Figures/Tubings/Functions/psic.pdf}}$.
\end{itemize}
Combining these results, 
we thus have
\begin{align}
	\psi_{\includegraphics[scale=0.6]{Figures/Tubings/Functions/psi+.pdf}} &= (X_1+X_2)^{\alpha_1+\alpha_2}\times\begin{cases}
 f_<(\xi) & |\xi|\le 1\,, \\[4pt] 	f_>(\xi) + h(\xi) & |\xi|\ge 1\,.
 \end{cases}\label{psi+series}
\end{align}
The solution for $\psi_{\includegraphics[scale=0.6]{Figures/Tubings/Functions/psi-.pdf}}$ can be obtained from this solution
 by the replacements $X_1,\alpha_1\leftrightarrow X_2,\alpha_2$.

\subsection{Factorization}

In Section~\ref{ssec:BC}, we argued that the general solution has $2^e$ free coefficients after imposing the absence of folded singularities for all basis functions. Indeed, we see in the present example that the solution 
has two free coefficients $c_{\includegraphics[scale=0.6]{Figures/Tubings/Functions/psi0.pdf}}$ and $c_{\includegraphics[scale=0.6]{Figures/Tubings/Functions/psic.pdf}}$, which appear in the total wavefunction 
\begin{align}
	\psi =\psi_{\includegraphics[scale=0.6]{Figures/Tubings/Functions/psi0.pdf}}+\big(\psi_{\includegraphics[scale=0.6]{Figures/Tubings/Functions/psi+.pdf}}+\psi_{\includegraphics[scale=0.6]{Figures/Tubings/Functions/psi-.pdf}}\big)\,.
	\label{equ:WF}
\end{align}
To obtain the full solution, we need to fix the relative normalization between these two coefficients. 
This can be done by taking the limit $X_1+Y\to 0$ and imposing the correct factorization behavior~\cite{Arkani-Hamed:2017fdk,Baumann:2020dch,Goodhew:2021oqg}:
\begin{align}
	\lim_{X_1+Y\to 0}\psi \propto (X_1+Y)^{\alpha_1}\Big((X_2-Y)^{\alpha_2}-(X_2+Y)^{\alpha_2} \Big)\,.
\end{align}
Substituting (\ref{equ:WF}) in the left-hand side, one finds
\begin{align}
	c_{\includegraphics[scale=0.6]{Figures/Tubings/Functions/psi0.pdf}} = -\frac{\Gamma(-\alpha_1)\Gamma(-\alpha_2)}{\Gamma(-\alpha_{1}-\alpha_2)}\,c_{\includegraphics[scale=0.6]{Figures/Tubings/Functions/psic.pdf}} \,.
\end{align}
The final solution for the wavefunction then is
\begin{align}
	\psi &= \,c_{\includegraphics[scale=0.6]{Figures/Tubings/Functions/psic.pdf}}\bigg[-\frac{\Gamma(-\alpha_1)\Gamma(-\alpha_2)}{\Gamma(-\alpha_{1}-\alpha_2)}(X_1+Y)^{\alpha_1}(X_2+Y)^{\alpha_2}\nn\\
	&\quad\quad +(X_1+X_2)^{\alpha_{1}+\alpha_2}\bigg(\frac{1}{\alpha_2}- \frac{\alpha_{1}+\alpha_2}{\alpha_1\alpha_2}\,{}_2F_1\left(1,\alpha_2,1-\alpha_1;-\frac{X_1-Y}{X_2+Y}  \right)+(1\leftrightarrow 2)\bigg)\bigg]\,,
\end{align}
where, for compactness, we have introduced the hypergeometric function.
For $\alpha_1=\alpha_2=\varepsilon$, this reproduces the result given in \cite{Arkani-Hamed:2023kig}.

%%%%%%%%%%%%%%%%%%%%%%%%%%%%%%%%%%%%%%%%%%%%%
\newpage
\section{Beyond Single Graphs}
\label{app:trp3}

The geometric patterns described in this paper were associated to individual Feynman graphs. It is natural to expect additional structure to appear in theories where physical observables combine together different graphs in a nontrivial way.
In~\cite{Arkani-Hamed:2023kig}, some of this additional structure was observed in ${\rm tr}\hs \phi^3$ theory of colored scalars when considering the differential equations for the sum over graphs (i.e.~inequivalent exchange channels). In particular, it was found that some of the basis functions are ``shared" between different channels. As we describe in this appendix, the techniques introduced in the main text can provide a more geometric interpretation of these patterns.
We will see that the geometries that organize the compatibilities between basis functions for a single graph and the geometry that organizes the compatibility between different exchange channels---the associahedron---combine in an interesting way.

\subsection{Four-Point Function}

We start with the case of the four-point function, which in $\phi^3$ theory arises from a two-site graph with a single exchange. The four external momenta form a closed quadrilateral (which, for simplicity, we will draw as a square). There are two possible
triangulations of this quadrilateral, corresponding to the $s$ and $t$-channel exchanges of the flavor-ordered wavefunction:
\beq 
\includegraphics[valign=c]{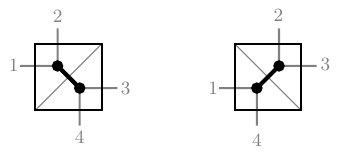}
\eeq
In the main text, we discussed the differential equations for a single graph, corresponding to one of these triangulations. Now, we consider the sum of the triangulations, $\psi \equiv \psi_s + \psi_t$.

\vskip 4pt
We focus on the connected part of the wavefunction. %, since adding in the disconnected parts is straightforward. 
 Both the $s$ and $t$-channel Feynman diagrams have a time-ordered and an anti-time-ordered contribution, which we assign to vertices of two line intervals. Taking derivatives with respect to the external energies collapses the internal propagators as before. Importantly, in all cases (i.e.~for both channels and both orderings in each channel) this leads to the {\it same} collapsed function. After collapsing the propagators, one cannot tell whether the contact solution came from the derivative of a function in the $s$ or $t$-channel.  We therefore assign the common source function to an edge connecting the four vertices describing the time-ordered functions in the two exchange channels:
\beq 
\raisebox{-1.3cm}{\includegraphics[scale = 1]{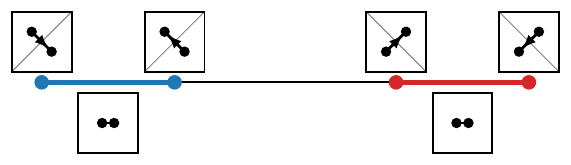}}
\label{equ:4pt-Assoc}
\eeq
In this picture, the functions involved in each exchange channel are assigned to line intervals as before, but now these intervals are connected by a line to illustrate the
shared source function.  
 This shared function reduces the number of basis functions (for the connected part) from the naive count of $2 \times 3 = 6$ to $5$.

\subsection{Five-Point Function}

As a more nontrivial example, we consider the five-point function, 
which in $\phi^3$ theory arises from a three-site graph. The kinematics is described by a pentagon, which has five triangulations, corresponding to five inequivalent exchange channels for the flavor-ordered wavefunction:
\beq 
\includegraphics[valign=c]{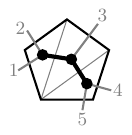}
\qquad \quad
\includegraphics[valign=c]{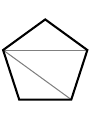}
\qquad \quad
\includegraphics[valign=c]{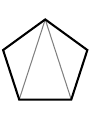}
\qquad \quad
\includegraphics[valign=c]{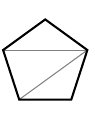}
\qquad \quad
\includegraphics[valign=c]{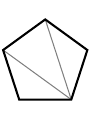}
\eeq
Both for amplitudes and the wavefunction, we can assign these triangulations to the vertices of the corresponding associahedron.
For the case of the five-point function, the associahedron is (by coincidence) also a pentagon.  The edges of the associahedron correspond to partial triangulations.

\vskip 4pt
We again focus on the connected part of the wavefunction. We have seen that it can be split into four functions that were assigned to the vertices of a square. To describe the compatibility of basis functions in the differential equations for the sum of graphs, we attach these squares to the vertices of the associahedron (one for each triangulation) and connect the edges:
\beq
\raisebox{-6cm}{\includegraphics[scale = 1.7]{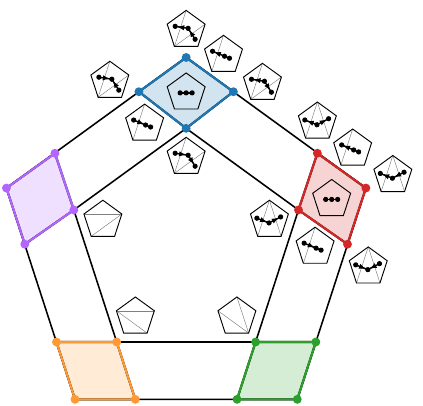}}
\label{equ:5pt-Assoc}
\eeq
We see that each edge of the pentagon connects two edges of these squares.
The functions associated to each edge are therefore shared between the different channels, meaning that they appears as a common sources in the differential equations for pairs of triangulations.  These connected intervals are the same as in~\eqref{equ:4pt-Assoc} for the four-point function, again reflecting the recursive structure of the equations.

\vskip 4pt
Similarly, the faces of all five squares describe the same function which is shared between all channels.  Instead of $5 \times 9 = 45$ basis functions (for the connected part of the wavefunction), the sum of graphs is described by a reduced basis with only $5\times 4+ 5\times 2 +1 = 31$  members. 

\vskip 4pt
Interestingly, the object in \eqref{equ:5pt-Assoc} can be viewed as two ``nested associahedra" (pentagons) with the edges of the inner pentagon extended until they intersect the outer pentagon. This automatically produces the five squares associated to the different exchange channels, as well as the shared functions between these channels.

\vskip4pt
The picture generalizes to higher points (corresponding to higher-dimensional geometries). For example, the basis functions of the six-point functions can be assigned to the vertices, edges, faces and the volume of a three-dimensional associahedron. Drawing two nested such associahedra and extending the faces of the inner associahedron produces the expected cubes at each of the 14 vertices. It also shows geometrically which faces and edges correspond to shared functions.

\vskip4pt
The geometric structure that organizes the sum over graphs is still somewhat mysterious. It tells us how to organize functions by associating functions to the various geometric components and allows us to read off the differential equations in essentially the same manner as for a single graph, but we still do not fully understand the origins of this structure, or whether it can be realized directly in some space of kinematics. Finally, it will be very interesting to understand the interplay between this geometry and the cosmohedron~\cite{Arkani-Hamed:2024jbp}.

\newpage
%%%%%%%%%%%%%%%%%%%%%%%%%
\phantomsection
%\enlargethispage{\baselineskip}
%\addtocontents{toc}{\protect\enlargethispage{\baselineskip}}
\addcontentsline{toc}{section}{References}
\bibliographystyle{utphys}
{\linespread{1.075}
	\bibliography{Flow-Refs}
}

\end{document}